\documentclass[useAMS,usenatbib]{mn2e}

 \usepackage{Times}

\usepackage{psfrag}
\usepackage{pspicture}
\usepackage{graphicx}

\newcommand\lsim{~\lower.5ex\hbox{$\buildrel < \over \sim$}~}
\newcommand\gsim{~\lower.5ex\hbox{$\buildrel > \over \sim$}~}

\def\gsim{~\lower.6ex\hbox{$\buildrel > \over \sim$}~}

\title[The evolution of $z=0.4-2.23$ SFGs]{The stellar mass function of star-forming galaxies and the mass-dependent SFR function since $\bf z=2.23$ from HiZELS}

\author[D. Sobral et al.]{David Sobral$^{1}$\thanks{Veni Fellow. E-mail: sobral@strw.leidenuniv.nl},  Philip N. Best$^{2}$, Ian Smail$^{3}$, Bahram Mobasher$^{4}$, John Stott$^{3}$, David Nisbet$^{2}$  \\
$^{1}$Leiden Observatory, Leiden University, P.O.\ Box 9513, NL-2300 RA Leiden, The Netherlands\\
$^{2}$SUPA, Institute for Astronomy, Royal Observatory of Edinburgh, Blackford Hill, Edinburgh, EH9 3HJ, UK\\
$^{3}$Institute for Computational Cosmology, Durham University, South Road, Durham, DH1 3LE, UK\\
$^{4}$University of California, 900 University Ave., Riverside, CA 92521, USA}
\begin{document}
\date{Accepted 2013 November 4. Received 2013 November 1; in original form 2013 September 13}

\pagerange{\pageref{firstpage}--\pageref{lastpage}} \pubyear{2013}
\maketitle

\label{firstpage}
\begin{abstract}

\noindent We explore a large uniformly selected sample of H$\alpha$ selected star-forming galaxies (SFGs) at $z=0.40,0.84,1.47,2.23$ to unveil the evolution of the star formation rate (SFR) function and the stellar mass function. We find strong evolution in the SFR function, with the typical SFR of SFGs declining exponentially in the last 11\,Gyrs as SFR$^*$($T$[Gyr])=10$^{4.23/T+0.37}$\,M$_{\odot}$\,yr$^{-1}$, but with no evolution in the faint-end slope, $\alpha\approx-1.6$. The stellar mass function of SFGs, however, reveals little evolution: $\alpha\approx-1.4$, M$^*\sim10^{11.2\pm0.2}$\,M$_{\odot}$ and just a slight increase of $\sim$\,2.3$\times$ in $\Phi^*$ from $z=2.23$ to $z=0.4$. The stellar mass density within SFGs has been roughly constant since $z=2.23$ at $\sim10^{7.65\pm0.08}$\,M$_{\odot}$\,Mpc$^{-3}$, comprising $\approx100$\% of the stellar mass density in all galaxies at $z=2.23$, and declining to $\approx20$\% by $z=0.40$, driven by the rise of the passive population. We find that SFGs with $\sim10^{10.0\pm0.2}$\,M$_{\odot}$ contribute most to the SFR density ($\rho_{\rm SFR}$) per d\,$\log_{10}$M, and that there is no significant evolution in the fractional contribution from SFGs of different masses to $\rho_{\rm SFR}$ or $\rho_{\rm SFR}$(d\,$\log_{10}$M)$^{-1}$ since $z=2.23$. Instead, we show that the decline of SFR$^*$ and of $\rho_{\rm SFR}$ are primarily driven by an exponential decline in SFRs at all masses. Our results have important implications not only on how SFGs need to be quenched across cosmic time, but also on the driver(s) of the exponential decline in SFR$^*$ from $\sim66$\,M$_{\odot}$\,yr$^{-1}$ to 5\,M$_{\odot}$\,yr$^{-1}$ since $z\sim2.23$.

\end{abstract}


\section{Introduction}\label{intro}

Our understanding of how galaxies form and evolve has increased dramatically over the last decades \citep[e.g. see reviews by][]{ELLIS,Robertson,Dunlop}. A wide range of surveys show that the star formation rate density, $\rho_{\rm SFR}$, rises to $z\sim2$ \citep[e.g.][]{Lilly96,Hopkins2006,Magnelli,Karim, Cucciati,Sobral13}, and reveal that the bulk of the stellar mass density seen in the Universe today was formed between $z\sim1-2$ \citep[e.g.][]{Marchesini,Sobral13,Muzzin13}.

Understanding why the Universe was so much more active in the past and which processes/mechanisms drive galaxy evolution are some of the most important open questions in the field of galaxy formation. The difficulty in assembling large, homogenous samples of galaxies spanning a wide redshift range has, however, been a strong limitation to our progress. Fortunately, the advent of large field-of-view cameras and multiplexing spectrographs on 4-8-m class telescopes (e.g. WFCAM, WIRCam, VISTA, FMOS, VIMOS) have made large area surveys (over-coming cosmic variance) based on a single, sensitive and well-calibrated selection over the full redshift range $0<z<3$ a reality. Such samples allow us to study evolutionary effects which otherwise can be simple consequences of biases associated with selection, small volumes and/or cosmic variance. 

Many studies have now highlighted some strong ``downsizing" trends. The term downsizing is arguably used in so many different contexts and to classify so many different results that it is often a very misleading term. It has been used to argue against the hierarchical model \citep[e.g.][]{Cimatti06,Fontanot,Cirasuolo10}, for example, but many claimed downsizing trends are a natural result of the hierarchical model \citep[e.g.][]{Neistein,Li08}. Some of the ``downsizing" trends have been used to argue that most of the activity in star-forming galaxies (SFGs) in the past happened in the most massive systems \citep{Cowie}, while now it happens mostly in galaxies with lower masses. Other studies \citep[e.g.][]{Juneau,Mobasher09,Simpson13} have revealed that the most massive galaxies had their peak of star formation around $z\sim2-3$, but contribute relatively very little to $\rho_{\rm SFR}$ below $z\sim2$. The downsizing term is also generally applied when describing the fact that low mass galaxies typically form more stars per unit mass (i.e., specific star-formation rate, sSFR) than more massive systems \citep[e.g.][]{Juneau,Zheng,Damen} and that the fraction of star-forming galaxies above a fixed SFR cut declines with mass \citep[e.g.][]{Brammer,Sobral11}. While studies reveal that these downsizing trends persist up to at least $z\sim1-2$, there is an expectation to see some changes beyond $z=2$ \citep[e.g.][]{Juneau}, unveiling an observational signature of the rough moment in time when even the most massive haloes/most massive galaxies at that time were still effective at forming stars and had not yet been quenched.

In addition, it is now known that the specific star formation rate (sSFR) of galaxies with the same mass increases with increasing redshift \citep[as $\sim(1+z)^{3}$ for M$\sim10^{10}$\,M$_{\odot}$ galaxies, e.g.][]{Koyama13}. Trends of increasing equivalent width of emission lines (a proxy for sSFRs) with redshift have also been found \citep[e.g.][]{Fumagalli}. The UV, H$\alpha$ and FIR luminosity functions also point towards a significant luminosity evolution ($L^*(z)$) consistent with the evolution of the sSFR, i.e., $\sim(1+z)^{3}$. The H$\alpha$ luminosity function evolution is found to be mostly driven by an increase of L$_{\rm H\alpha}^*$ up to $z=2.23$ \citep[c.f.][]{Sobral13,Stott,Colbert}, while its faint end slope, $\alpha$, is found to be constant at $\alpha=-1.6$ \citep{Sobral13}. These results are in good agreement with the UV LFs  \citep[c.f.][]{Smit,Alavi13}. Observations are therefore pointing towards the evolution in star formation rates (SFRs) or sSFRs being the most important feature of the evolution of star-forming galaxies \citep[c.f.][]{Peng,Peng12}.

Determining the mass (amount of stars already formed) and SFR (amount of stars forming) functions of galaxies, and their evolution across time, is of fundamental importance to improve our understanding of how galaxies form and evolve, and address many of the outstanding questions. Over the past years, a significant effort has been put into determining these, although often with different approaches/selections and separate analyses. Studies either focus on a ``star-forming" population (SFR or sSFR selected) or a ``passive" population (continuum-selected). A remarkable advance has been obtained in the determination of the stellar mass function and its evolution since $z\sim3-4$. The latest studies \citep[e.g.][]{Marchesini,Peng,Ilbert,Muzzin13} suggest that both M$^*$ and $\alpha$ evolve very little since $z\sim2$, while the normalisation, $\Phi^*$, is the major parameter evolving in the 11 billion years since then. Recent studies have also started to investigate the star-formation rate function (SFR function) and its evolution, although this is often a more complicated function to determine (when compared to the mass function), due to the selection and difficulty of converting SF indicators to SFRs after taking into account the effects of dust and star-formation timescales. Nonetheless, the large statistical samples, coupled with robust and statistical dust corrections \cite[e.g.][]{GarnBest} that have now been tested beyond the local Universe \citep[e.g.][]{Sobral12,Dominguez,Ibar13,Price}, are starting to allow us to compute them. \cite{Smit} showed the evolution of the SFR for $z>3$, and other studies \citep{Martin,Bothwell} have derived it for the local Universe.

So far, no study has robustly determined both the SFR and mass functions for star-forming galaxies at $0<z<3$ (during which the vast majority of the stellar mass density was assembled), nor evaluated the contribution to $\rho_{\rm SFR}$ from star-forming galaxies with different masses since $z\sim3$. Here we will overcome previous short-comings and limitations, by using the largest homogeneous samples of H$\alpha$ selected star-forming galaxies \citep{Sobral13} at four different redshifts ($z=0.40,0.84,1.47,2.23$), covering the peak and fall of the $\rho_{\rm SFR}$. This paper is organised in the following way: \S 2 presents the sample, stellar masses, SFRs and sSFRs, while \S 3 presents the methods and procedures adopted to derive SFR and stellar mass functions. \S 4 shows the results: the SFR function, the stellar mass function of star-forming galaxies and their evolution and SFR functions for samples with different masses, quantifying the contribution from different masses to the cosmic star formation history. Finally, we provide our conclusions in \S 5. An H$_0=70$\,km\,s$^{-1}$\,Mpc$^{-1}$, $\Omega_M=0.3$ and $\Omega_{\Lambda}=0.7$ cosmology is used and, except where otherwise noted, magnitudes are presented in the AB system. Throughout this paper we use a Chabrier IMF to obtain both stellar masses and SFRs; using a Salpeter IMF would lead to systematically higher stellar masses (including M$^*$) and SFRs (including SFR$^*$), but the relative SFRs and stellar masses would remain unchanged, and thus the overall results of this paper would not change.

\section{THE SAMPLES OF STAR-FORMING GALAXIES}\label{data_technique}

\subsection{The HiZELS survey}\label{HIZELS}

The High Redshift Emission Line Survey \citep[HiZELS;][]{G08,S09a,S09b,Sobral12,Sobral13,Best2010} is a Campaign Project using the Wide Field CAMera (WFCAM, \citealt{Casali}) on the United Kingdom Infra-Red Telescope (UKIRT) which exploits custom designed narrow-band filters in the $J$ and $H$ bands (NB$_J$ and NB$_H$), along with the H$_{2}$S1 filter in the $K$ band, to undertake large-area, moderate depth surveys for line emitters. HiZELS targets the $\rm H{\alpha}$ emission line redshifted into the near-infrared wavelengths at $z = 0.84, 1.47 \rm \,and \,2.23$ using the three different filters. The survey is fully complemented by deeper narrow-band observations with Subaru Suprime-Cam NB921 imaging \citep{Sobral12,Sobral13} to obtain $\rm H{\alpha}$ emitting galaxies at $z=0.4$ and the [O{\sc ii}]\,3727 emission from the $z=1.47$ $\rm H{\alpha}$ sample \citep[see also][]{Hayashi13}, as well as deeper WFCAM and Very Large Telescope near-infrared imaging through the H$_{2}$S1 filter in selected fields \citep{Sobral13}. The survey was designed to trace star-formation activity across the likely peak of the star formation rate density in the Universe and provide detailed information about a well-defined statistical sample of star-forming galaxies at each epoch (see \citealt{Best2010}, but also e.g. \citealt{Swinbank12a,Swinbank12b,Sobral13b}). HiZELS provides uniformly selected, large samples ($\sim1000$ per redshift slice) of H$\alpha$ emitters covering a very wide range of environments and properties, and it is therefore ideal for the purposes of this paper.

%
%
\begin{figure*}
\begin{minipage}[b]{0.32\linewidth}
\centering
\includegraphics[width=5.7cm]{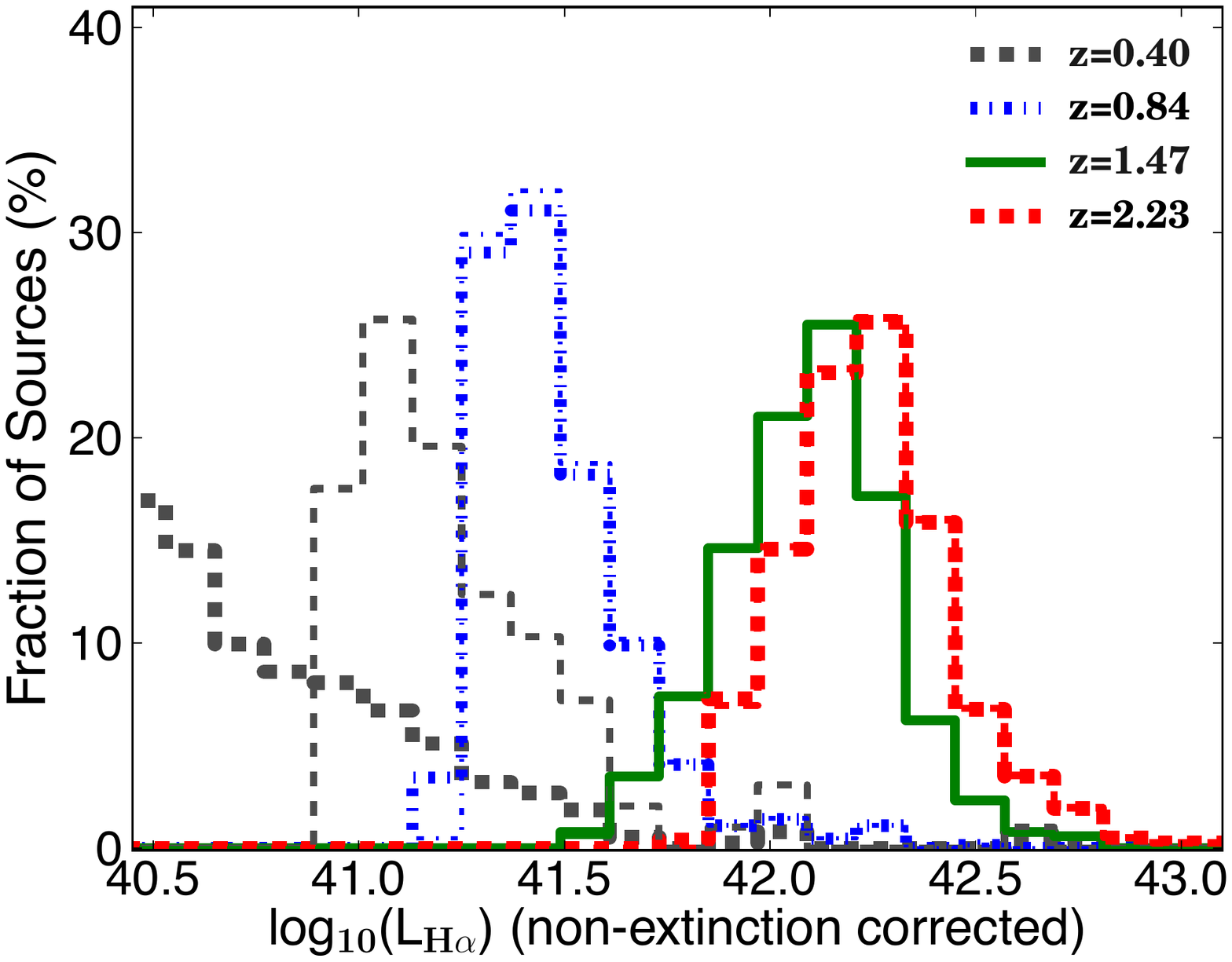}
\end{minipage}
\hspace{0.1cm}
\begin{minipage}[b]{0.32\linewidth}
\centering
\includegraphics[width=5.7cm]{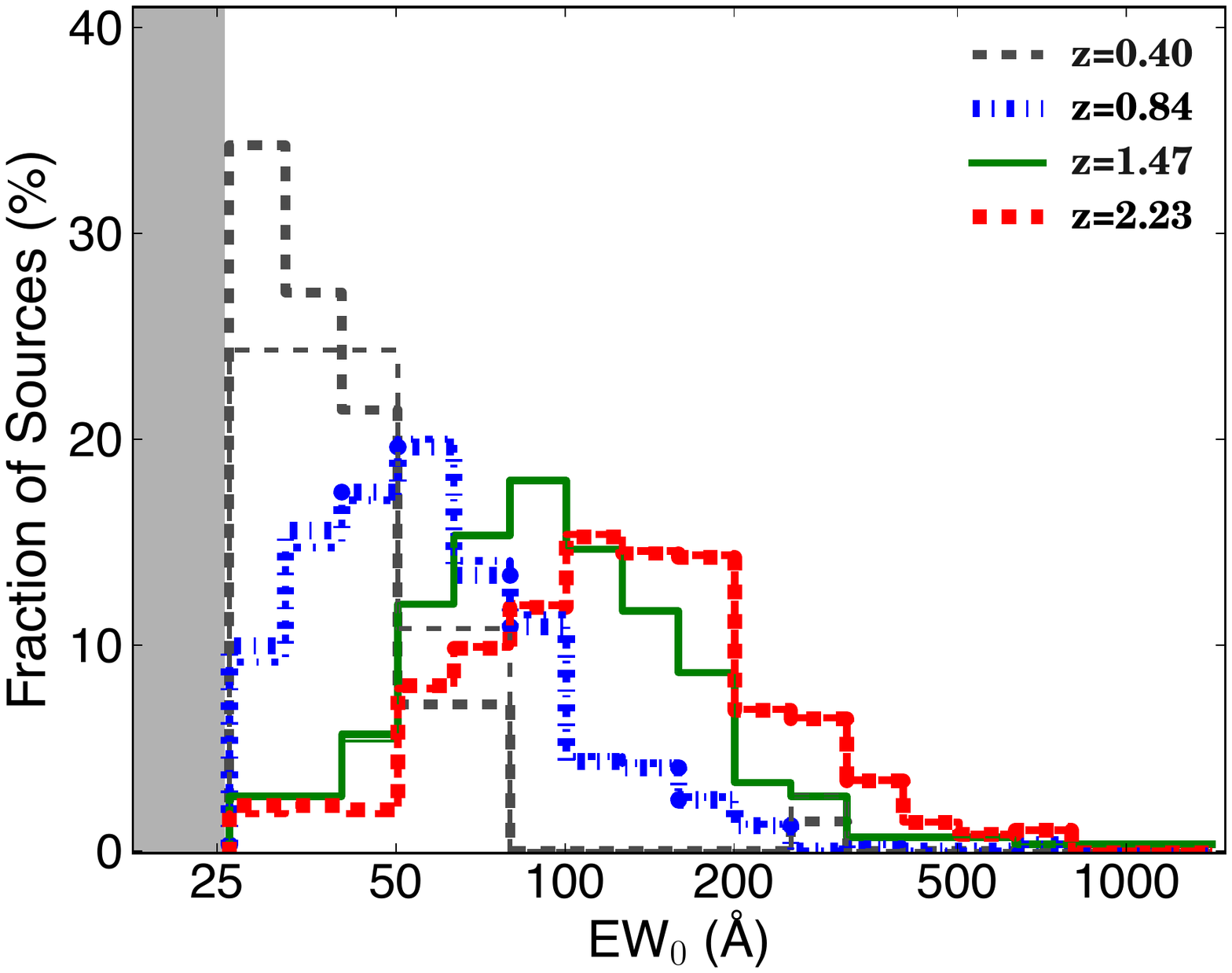}
\end{minipage}
\hspace{0.1cm}
\begin{minipage}[b]{0.33\linewidth}
\centering
\includegraphics[width=5.7cm]{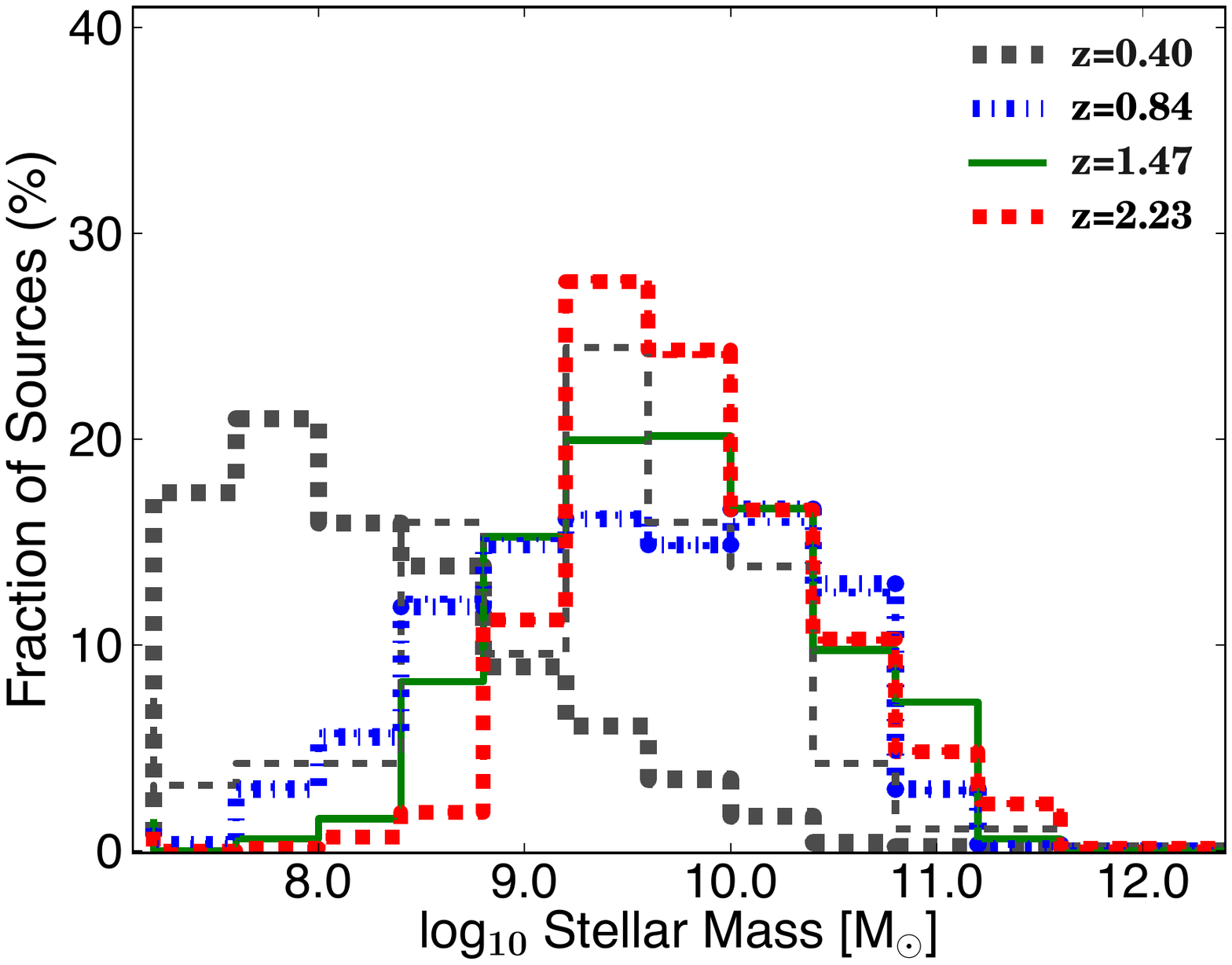}
\end{minipage}
\caption[EWs]{{\it Left}: The normalised distribution (thicker lines) of observed H$\alpha$ luminosities (without dust correction). In order to provide a fairer comparison between the $z=0.4$ sample and those at higher redshift, we also show the 0.25\,SFR$^*$ cut in a thinner line than the full sample. {\it Middle}: The distribution of rest-frame equivalent widths for the entire H$\alpha$ samples at the different redshifts (for stellar masses higher than 10$^{9.0}$\,M$_{\odot}$). The shaded region highlights the common EW selection (down to the same rest-frame equivalent width, 25\,\AA), and shows where the sample is incomplete in EW. We show the 0.25\,SFR$^*$ cut in a thinner line than the full sample. Regardless of the cut, there is a clear evolution in the EWs. {\it Right}: The normalised distribution of stellar masses of all H$\alpha$ emitters at the different redshifts and after applying the 0.25\,SFR$^*$ cut (shown in a thinner line). Note that both the volumes and H$\alpha$ luminosity limits vary, and thus without any cut the distribution of masses at $z\sim1-2$ is particularly different from $z=0.4$. A simple SFR$^*$ cut that guarantees that the data-sets are relatively complete down to that limit results in a common similar normalised mass distribution. The mass function derived later in this paper takes into account all sources of incompleteness and the effects of the flux and EW selection. \label{HISTOGRAMS}}
\end{figure*}

\subsection{The sample of star-forming galaxies at $\bf z\sim0.4-2.23$}\label{sample}

We use the large HiZELS sample of $z = 0.4, 0.84, 1.47 \rm \,and \,2.23$ $\rm H\alpha$-selected emitters in both the UKIRT Infrared Deep Sky Survey, Ultra Deep Survey (UKIDSS UDS, \citealt{Lawrence}, Almaini et al. in preparation) and The Cosmic Evolution Survey (COSMOS, \citealt{Scoville,Capak}) fields as described in \cite{Sobral13}. We refer the reader to that paper for full details of the catalogues used. These data cover a typical area of $\sim2$\,deg$^2$ (UDS+COSMOS) with each narrow-band filter (the exact coverage depends on the field and waveband). The narrow-band excess sources are visually inspected to remove image artefacts. We use spectroscopic redshifts, double/triple line detections, high-quality photometric redshifts and optimised colour-colour selection \citep[see][]{Sobral13} to yield large and pure samples of H$\alpha$ emitters (see Table \ref{numbers}). All the samples have been homogeneously selected down to the same rest-frame $\rm H\alpha$+[N{\sc ii}] equivalent width lower limit of 25\,\AA \ (see Figure \ref{HISTOGRAMS}). The $z=0.4$ sample reaches down to much lower H$\alpha$ luminosities than the others, but it is possible to obtain fully comparable samples by applying a limit in SFR$^*$($z$) or L$^*$($z$) of $>0.25$SFR$^*$; this takes into account the cosmic evolution of that parameter. Table \ref{numbers} presents the number of sources if such cut is applied. We use this sample for the completeness analysis and to compare samples in Figures 1 and 3. We check that our results are fully recovered with this sample, although they are naturally affected by significantly larger errors at $z=0.4$ (mostly due to the reduction of the sample size by a factor $\sim10$ at $z=0.4$) and/or probe a much narrower parameter space in SFR and stellar mass. We therefore chose to use the full sample for the rest of the analysis, except when noted otherwise.

%
%
\begin{table}
 \centering
  \caption{A summary of the number of H$\alpha$ emitters at the different redshifts for the full sample \citep[see][]{Sobral13} and if one applies a cut which take into account the cosmic evolution of SFR$^*$($z$) or L$^*$($z$). The sample with a cut at $>0.25$ SFR$^*$ is used in Figures 1 and 2 and to check the validity of our results, but the full sample is used for the rest of the paper as that provides much better statistics at $z=0.4$.}
  \begin{tabular}{@{}ccccc@{}}
  \hline
  Sample of H$\alpha$ SFGs  & $z=0.40$ & $z=0.84$ & $z=1.47$ & $z=2.23$  \\
 \hline
   \noalign{\smallskip}
Entire Sample & 1123 & 637 & 515 & 772 \\
$>0.25$ SFR$^*$($z$)  & 97 & 618 & 514 & 766 \\
 \hline
\end{tabular}
\label{numbers}
\end{table}

\subsection{Stellar masses for H$\alpha$ emitters}\label{masses}

Stellar masses are obtained by spectral energy distribution (SED) fitting of stellar population synthesis models to the rest-frame UV, optical, near- and mid-infrared data available ($FUV, NUV, U, B, g, V, R, i, I, z, Y, J, H, K, 3.6\mu \rm m, 4.5\mu \rm m,$ 5.8$\mu \rm m$, 8.0$\mu \rm m$), following \cite{Sobral11}. The SED templates are generated with the \cite{BC03} package using \cite{B07} models, a \cite{Chabrier} IMF, and exponentially declining star formation histories with the form $e^{-t/\tau}$, with $\tau$ in the range 0.1 Gyrs to 10 Gyrs. The SEDs were generated for a logarithmic grid of 200 ages (from 0.1 Myr to the maximum age at each redshift). Dust extinction was applied to the templates using the \cite{Calzetti} law with $E(B-V)$ in the range 0 to 0.5 (in steps of 0.05), roughly corresponding to A$_{\rm H\alpha}\sim0-2$. The models are generated with five different metallicities ($Z=0.0001-0.05$), including solar ($Z=0.02$). We use the best-fit template to obtain one estimate of the stellar mass, but we also compute the median stellar mass across all solutions in the entire multi-dimensional parameter space for each source, which lie within 1$\sigma$ of the best-fit. The two estimates correlate very well, but, as expected, the best-fit mass is very sensitive to small changes in the parameter space and/or error estimations in the data-set, while the median mass of the 1$\sigma$ best fits is robust against such variations. Thus, throughout this paper, we use the median mass of the 1$\sigma$ best-fits, instead of the best-fit SED mass, but we have also checked that all our results and conclusions remain unchanged if we use stellar masses which result from the best-fit SED.

We show the distribution of stellar masses for the samples of H$\alpha$ emitters at the four different redshifts in Figure \ref{HISTOGRAMS}. H$\alpha$ emitters in our $z=0.84-2.23$ samples as a whole have a typical mass of $\sim10^{9.75}$\,M$_{\odot}$, although they are found to have a range of masses $\sim10^{8.5}-10^{11.5}$\,M$_{\odot}$. The $z=0.40$ sample is much deeper (by an order of magnitude), whilst covering only a fraction of the volume, and thus it is sensitive to star-forming galaxies down to much lower masses (see Figure \ref{HISTOGRAMS}). As Figure\ref{HISTOGRAMS} shows, applying the $>0.25$ SFR$^*$ cut (to make the $z=0.4$ sample comparable to those at higher redshifts, see \S\ref{sample}) leads to a $z=0.4$ stellar mass distribution which is much more similar to that of the $z=0.84,1.47,2.23$ samples.

\subsubsection{The effect of using complex SFHs}

A potential concern about fitting the SEDs of our H$\alpha$ star-forming galaxies with a single exponential model for the star-formation history is that this could lead to errors or biases in the estimation of the stellar masses. This is because it is likely that, especially at the lower redshifts, many of the H$\alpha$-selected galaxies may possess a younger stellar population or recent starburst (leading to the selection as an H$\alpha$ emitter) on top of an underlying older stellar population.  In this case the young stellar population would dominate the SED in the optical waveband, strongly influencing the SED fitting, but the older stellar population may contain the bulk of the stellar mass. To test the robustness of the stellar mass determinations, we compared SED fits to our $z=0.84$, $z=1.47$ and $z=2.23$ galaxies (the samples obtained with the original HiZELS narrow-band filters) using both a single exponential star-forming history, and an exponential model with an additional recent starburst. The young stellar population was modelled as a 30\,Myr top-hat burst of star-formation (this value chosen as being the approximate star-formation timescale traced by H$\alpha$), at solar metallicity. The dust attenuation of this component was fitted independently, using a Calzetti law with E(B-V) ranging up to 0.6. The relative contribution of the young starburst was allowed to vary between 0 and 30\% of the stellar mass of the galaxy.

\begin{figure}
\centering
\includegraphics[width=8.2cm]{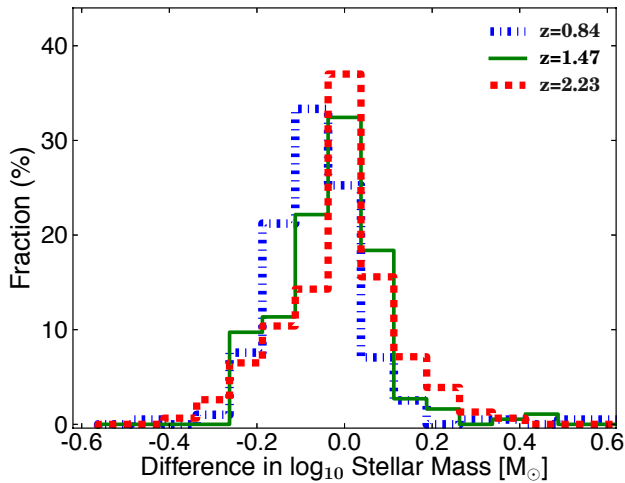}
\caption[Mass function]{Histograms of the difference in estimates of stellar mass derived with compound star-formation histories models and those with simple models (mass with compound star-formation histories $-$ mass with simple models). These show that the stellar mass estimates obtained with a single stellar population and simple exponential star formation histories are in very good agreement with measurements including more complex star formation histories. The scatter in the differences is smaller than the expected individual uncertainties in mass.  \label{mass_diff}}
\end{figure}

These compound model fits do not cover all of the possible parameter space for the star-formation history of the galaxies, but they do allow a direct comparison of very different possibilities, to indicate the robustness of the SED parameter estimates from single exponential models. We find that in the compound models the ages of the old stellar population are typically 50-200\% larger than the age determined in the simple model, and the old stellar population is often fitted with a lower metallicity model. The current star-formation rates resulting from using compound models are higher by a factor of $\approx2$, and better match the H$\alpha$ estimates. However, as Figure \ref{mass_diff} shows, the stellar masses do not change dramatically between the simple and compound models, being reduced by only 0.06, 0.03 and 0.00 dex (compared to the single exponential model) at $z=0.84$, $1.47$ and $2.23$ respectively, with a scatter in each case of $\sim0.12$\,dex.

%
%

We conclude that, at least for datasets including wide wavelength coverage through to the rest-frame near-IR, the use of single exponential star-formation models does not introduce any significant biases into the derivation of stellar masses. However, we caution against the use of all other SED-derived properties (such as age, metallicity, dust extinction, current SFR) for star-forming galaxies, without detailed modelling of the star-formation history.

\subsection{EWs, SFRs and sSFRs}\label{EWs_SFRs}

We compute observed equivalent widths (EW) of H$\alpha$+[N{\sc ii}] (all narrow-band filters are wide enough to encompass both H$\alpha$ and the adjacent [N{\sc ii}] lines) using:
\begin{equation}
   {\rm EW}_{{\rm observed}}=\Delta\lambda_{NB}\frac{f_{NB}-f_{BB}}{f_{BB}-f_{NB}(\Delta\lambda_{NB}/\Delta\lambda_{BB})},
\end{equation}
where $\Delta\lambda_{NB}$ and $\Delta\lambda_{BB}$ are the FWHMs of the narrow- and broad-band filters \citep[see][]{Sobral13}, and $f_{NB}$ and $f_{BB}$ are the flux densities measured for the narrow and broad-bands, respectively. Rest-frame EW (EW$_0$) are computed as EW$_{\rm observed}$/$(1+z)$. We show the EW$_0$ distribution of the entire sample in Figure \ref{HISTOGRAMS}. For simplicity we refer to EW$_0$(H$\alpha$+[N{\sc ii}]) as EW$_{\rm H\alpha+[N{\sc II}]}$.

H$\alpha$ fluxes are obtained by first computing the emission line flux within the narrow-band filter \citep[see][]{Sobral13}, which contains some contribution from the adjacent [N{\sc ii}] line, and then removing the contribution from the [N{\sc ii}] line. This contribution is estimated and removed using the relation between metallicity (the [N{\sc ii}]/H$\alpha$ line ratio) and EW$_{\rm H\alpha+[N{\sc II}]}$ from SDSS \citep[][]{Villar08,Sobral12}. This is a source of potential uncertainty, particularly on a source by source basis, but we note that we apply the same correction for the samples at all redshifts and that our bins are large enough to eliminate the bulk of the source-by-source variations; therefore any trends with redshift and EW$_0$ or sSFR are independent of this correction. We also note that such relation seems to hold very well at both $z=0.84$ and $z=1.47$ \citep[][]{Stott13b,Sobral13b}. H$\alpha$ fluxes are converted to H$\alpha$ luminosities for each redshift slice. The distributions of observed H$\alpha$ luminosities (after removing the contribution from [N{\sc ii}], but without applying any dust correction) are shown in Figure \ref{HISTOGRAMS} for the samples at the different redshifts.


The H$\alpha$ luminosities are based on 2$''$ diameter aperture photometry for $z=0.8,1.47,2.23$ and 3$''$ diameter aperture photometry for $z=0.4$, in order to select and measure the H$\alpha$ line over $\sim16$\,kpc (diameter) at all redshifts. While such apertures are expected to recover the bulk of the H$\alpha$ flux, they can miss a fraction of the total flux, due to a combination of extended H$\alpha$ emission, and seeing. We investigate this by stacking NB-BB (line emission) $15\times15''$ thumbnails H$\alpha$ emitters and comparing their 2$''$ (or 3$''$, for $z=0.4$) flux with the total stacked flux. We find that we miss 23$\pm2$ per cent of the total flux in the 2$''$ apertures ($z=0.84$,1.47,2.23), and 12$\pm3$ per cent of the flux in $3''$ apertures ($z=0.40$). We split the sample in bins of luminosity and mass to test for any strong dependence on the missing flux with such properties. We do not find any significant correlation, as the variations are always smaller than the typical errors ($\sim5$ per cent). We therefore apply an aperture correction of 1.3 to the 2$''$ measurements and a correction of 1.14 to the 3$''$ measurements. These corrections are relatively small and we note that they do not change any of the results in this paper. We then compute star formation rates from the aperture corrected $\rm H{\alpha}$ luminosities using the relation from \cite{Kennicutt}, corrected for a Chabrier (2003) IMF:
\begin{equation}
{\rm SFR [M_{\odot}yr^{-1}]}=4.4\times10^{-42}\rm L(H\alpha) \rm[erg \,s^{-1}].  
\end{equation}
In \cite{Sobral13}, a constant dust extinction $A_{H{\alpha}}=1$\,mag was used to correct H$\alpha$ luminosities. Here we use a more sophisticated correction to obtain extinction corrected SFRs. We use the robust empirical relation between median stellar mass and median dust extinction determined by \cite{GarnBest}. We note that whilst the relation has been derived for a large SDSS sample, it has been shown to hold up to at least $z\sim1.5$ by \cite{Sobral12} \citep[and further confirmed by][]{Dominguez,Ibar13}, with the same slope and normalisation. This contrasts with other statistical relations (e.g. SFR-observed H$\alpha$ luminosity), which are clearly shown to evolve with redshift \citep[e.g.][]{Sobral12,Dominguez}, and thus not valid for different cosmic epochs. We also note that the main conclusions of this paper do not depend on the extinction correction adopted, and that they are still recovered if a homogeneous/simple dust extinction correction is used, or if no dust extinction correction is applied. Finally, specific star formation rates (sSFR) are computed by obtaining the ratio between our dust-corrected H$\alpha$ SFR and stellar mass for each individual galaxy.

\section{METHODS: DETERMINING SFR and MASS FUNCTIONS}\label{methods}

Here we present how we derive SFR and stellar mass functions for our sample of H$\alpha$ star-forming galaxies at $z=0.4-2.23$ and describe our corrections for completeness due to the observational limits in the samples: H$\alpha$ observed flux and H$\alpha$ EW. Our samples are all selected down to the same rest-frame EW (see Figure \ref{HISTOGRAMS}), so measuring the evolution of H$\alpha$ EW and its potential dependence on stellar mass is of key importance to derive any necessary corrections when computing the SFR and stellar mass functions for star-forming galaxies. 

\subsection{Stellar Mass-EW$_{\rm H\alpha}$ dependence: completeness}  \label{EW_MASS}

\begin{figure}
\centering
\includegraphics[width=8.2cm]{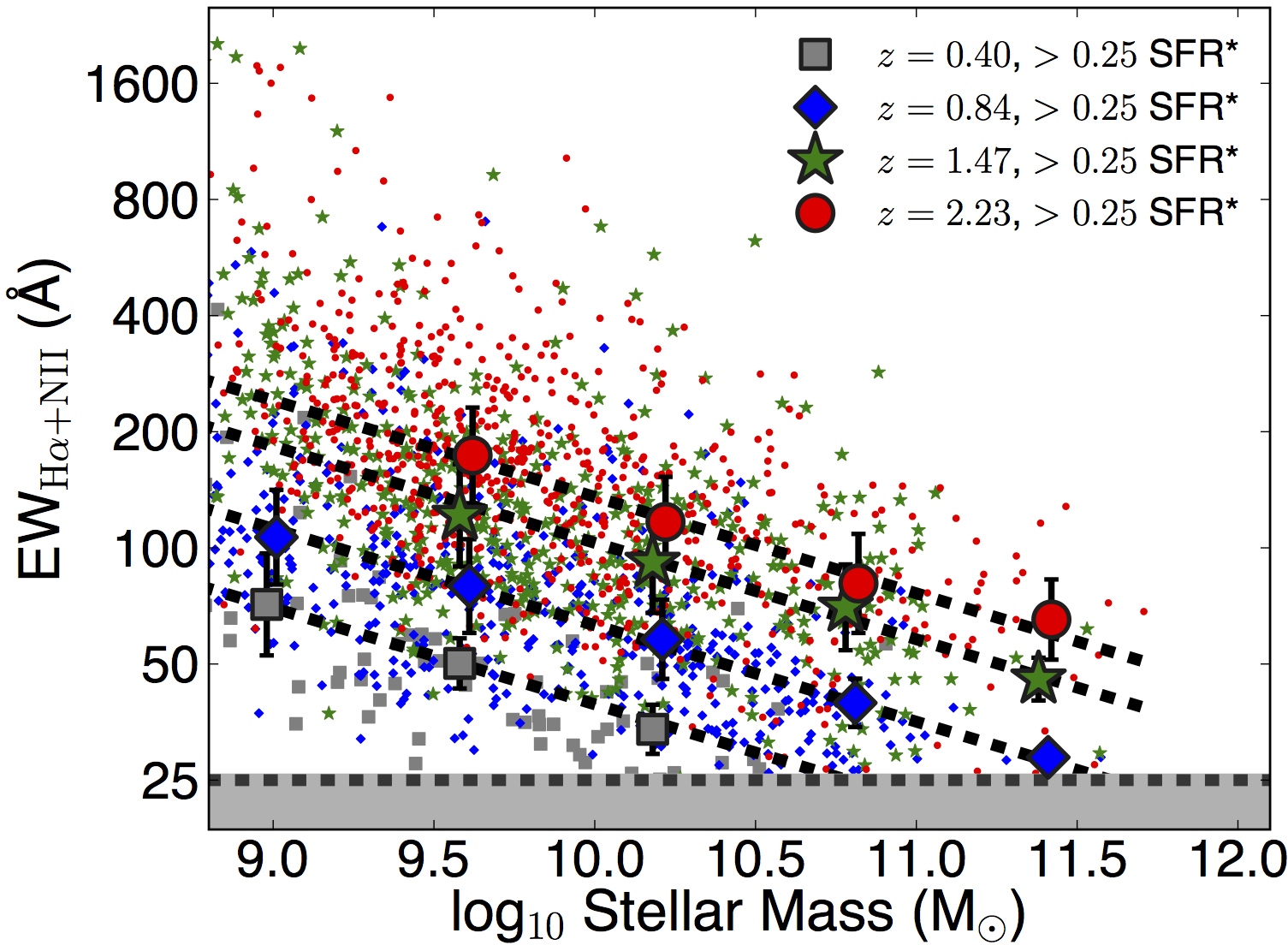}
\caption[EWs matched]{The relation between rest-frame EW (and median EW and error on the median) with mass for the four different redshifts. We apply a SFR$^*$ cut to make the $z=0.4$ sample more comparable to the higher redshift samples. We apply small offsets of $\pm0.02$\,dex in stellar mass for the median EW points for presentation purposes. The median EW is found to decline steeply with mass at all times with roughly the same slope and to decline with decreasing $z$ at all masses. We find a relation consistent with a constant slope (non-evolving with redshift) given by $\rm EW(M)\sim M^{-0.25}$ for a fixed redshift, and EW$(z)\sim(1+z)^{1.72\pm0.06}$ for a fixed mass.  \label{EWvsMASS}}
\end{figure}

Figure \ref{EWvsMASS} shows the evolution of the (rest-frame) EW (H$\alpha$+[N{\sc ii}]) with redshift and with mass for the entire sample, down to a stellar mass of 10$^{9}$\,M$_{\odot}$, a rough common mass completeness of all our samples (at $z>1.0$ the completeness is actually closer to 10$^{9.5}$\,M$_{\odot}$). Here we apply a $>0.25$SFR$^*$ cut in order to make the samples at $z>0.8$ and $z=0.4$ comparable, and thus allowing us to study the evolution of the EW vs stellar mass. We apply that cut here because the $z=0.4$ sample is significantly deeper and thus includes a much larger number of galaxies forming stars at a rate significantly below the average. This would bias the comparison between $z=0.4$ and the other higher redshift samples, particularly at low masses. Above our common mass completeness, the decline in median EW is well fitted by a single slope at all redshifts, given by EW=M$^{-0.25\pm0.01}$, with the normalisation evolving as $(1+z)^{1.72\pm0.06}$ -- in very good agreement with the literature \citep[e.g.][]{Fumagalli}.

%
%

Figure \ref{EWvsMASS} also highlights our selection limit in EW. NB surveys require an EW cut, so Figure \ref{EWvsMASS} is also extremely helpful in accessing the potential incompleteness of other samples selected down to different EWs and at different redshifts. Down to our EW limit (H$\alpha$ + [N{\sc ii}], 25\,\AA), we find that the samples at $z\sim1-2.2$ are relatively complete to even the highest masses, but that at $z<0.8$ even this EW limit will lead to be biased towards lower mass star-forming galaxies. As the EW continues to decline with declining redshift, samples at $z<0.4$ will only be complete up to relatively low mass galaxies (more massive systems will only make it into the samples if they have very high EWs/sSFRs -- and these are rare). 

We note that our samples are SFR (H$\alpha$ flux) and EW limited, but that, mostly because of the EW limit, the sample can be incomplete in mass, particularly at the highest masses and at the lowest redshifts (see Figure \ref{EWvsMASS}). We use the results from Figure \ref{EWvsMASS} to estimate the necessary completeness corrections which are particularly important for the $z=0.4$ sample at masses $>10^{10.5}$\,M$_{\odot}$, where the EW cut results in a significant incompleteness. We do this by using the $z=1.47$ and $z=2.23$ samples and evaluate the fraction of massive galaxies that would be missed if the difference between the EW limit and the median EW of $\sim10^{10}$\,M$_{\odot}$ galaxies at $z\sim2$ were to be the same as it is at $z=0.4$. We find that the two mass bins above $10^{10.5}$\,M$_{\odot}$ at $z=0.4$, should be corrected by a factor of 2.3. As this correction is relatively uncertain, we add 30\% of the correction in quadrature to the final errors. We also follow this procedure for the $z=0.84$ sample, although the corrections factor in this case is only 1.2, half of that at $z=0.4$. Furthermore, in order to minimise the errors and so guarantee that our approach is fully valid (using completeness and volume corrections derived from our selection function), we use relatively broad bins of 0.3\,dex in mass. These are larger than, or at least comparable to the errors in our masses, which vary from $0.1-0.25$\,dex (depending on redshift and mass).

\subsection{SFR Function}  \label{SFR_FUNCT_METHOD}

We compute the star formation rate (SFR) function of the Universe and its evolution with redshift since $z=2.23$. We estimate the SFR functions using the same method described in \cite{Sobral13}, but adapting it to reflect the more sophisticated dust corrections (as a function of mass). In summary, the simulations from \cite{Sobral13} are used to both obtain completeness and filter profile corrections. While H$\alpha$ luminosities are easily linked with observed fluxes and EWs (which are the two selection criteria), here SFRs are corrected for extinction based on stellar mass, so the relation contains extra scatter. In order to avoid potential problems introduced by such scatter, our final incompleteness corrections are obtained using the observed EWs and H$\alpha$ fluxes of all the sources in each SFR bin.

\subsection{Mass Function}  \label{SFR_FUNCT_METHOD}

We construct the mass functions for star-forming galaxies and evaluate them for $z\lsim2.23$, using a homogeneous sample of H$\alpha$ selected galaxies spanning the redshift range $0.4 \lsim z \lsim2.23$. This is not the same as selecting galaxies in a (typically redshift-dependent) rest-frame continuum band which is often erroneously referred to as ``mass-selected". Our approach is arguably much cleaner and results in a well-understood sample of star-forming galaxies. Also, because these are star-forming galaxies, the mass-to-light ratios result in our samples being complete to much lower masses than for surveys looking at passive/more general population of galaxies \citep[where mass-to-light ratios are very high, and thus they are only complete to significantly higher masses, e.g.][]{Muzzin13}.

In order to apply appropriate completeness corrections, we use the simulations from \cite{Sobral13} which provide the completeness of each galaxy as a function of H$\alpha$ flux/luminosity and that take into account the exact selection function that was used to obtain the samples (including the common EW$_0$ limit and the differences in depth across fields and sub-fields). We use the estimated completeness corrections (based on observed H$\alpha$ flux) of each source to weight the number of star-forming galaxies that are likely to be in a mass bin if the sample was 100\% complete at that flux.


We note that our measurements and those usually presented in the literature use a \cite{Chabrier} IMF. For comparison, a Salpeter IMF will result in masses which are on average a factor 1.6 higher. This would result in an increase in M$^*$, but our conclusions would remain completely unchanged (as M$^*$ would increase for all mass functions and the stellar mass density also increases for all samples and redshifts).

\subsection{SFR Functions for different masses}  \label{SFR_FUNCT_M_METHOD}

We also divide the sample of SFGs at each redshift into different stellar mass bins in order to evaluate the dependence of the SFR function and its integral ($\rho_{\rm SFR}$) on stellar mass. We repeat the process of deriving SFR functions for the sub-samples of galaxies in each mass bin and evaluate $\rho_{\rm SFR}$(M). We note that because our dust-corrections are mass-dependent, the single H$\alpha$ flux limit of our sample will result in different SFR limits for the different sub-samples, being higher for sub-samples with higher masses. In order to avoid any biases due to this effect, we only compute the SFR function down to our completeness limit after taking this into account.

We fit Schechter functions to the SFR functions for different mass bins and make use of the integral of the mass functions in those mass bins to further constrain the fits. In practice, we require that the number density of the fully integrated SFR function fit for any given mass bin (and for each redshift) is within $\pm0.3$\,dex (the typical errors) of the number density of galaxies within that range implied by the mass function. This allows us to better constrain possible values of $\alpha$, but mostly avoids unphysical combinations of parameters as best-fits which are strongly disfavoured by the mass function. We note that a Schechter function may not be the most appropriate form to fit to these mass-dependent SFR functions \citep[][]{Salim2012}, and therefore the values of $\alpha$ should be taken with caution. Nevertheless, a Schechter function provides a good, simple fit to the data, and because we constrain the parameters by using the appropriate integral of the mass function, the integral of our SFR function ($\rho_{\rm SFR}$) is relatively unaffected by the change of $\alpha$ or functional form.

\section{RESULTS} \label{RESULTS}

%
%
%
%
\begin{figure*}
\begin{minipage}[b]{0.49\linewidth}
\centering
\includegraphics[width=8.7cm, height=6.63cm]{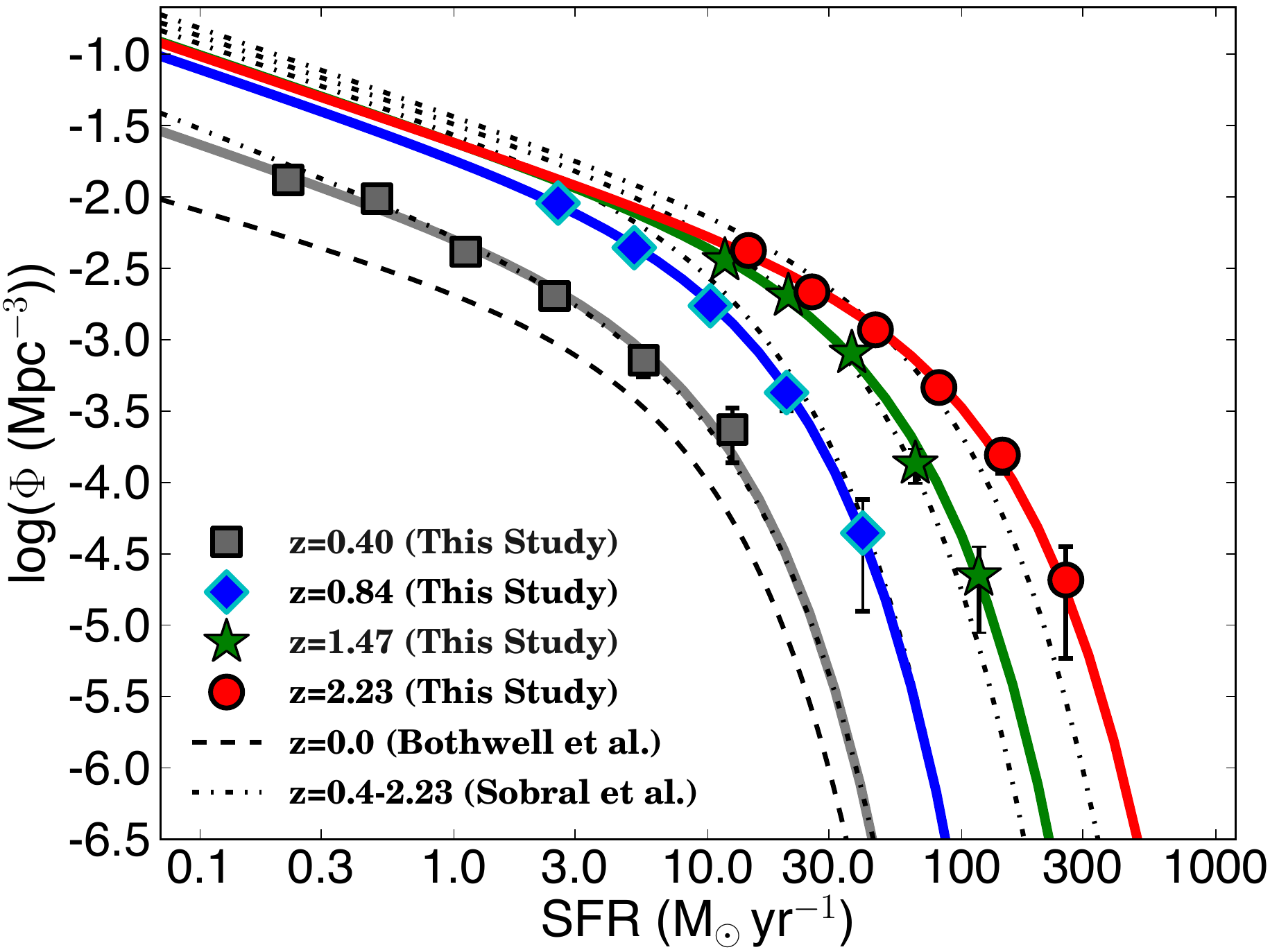}
\end{minipage}
\hspace{0.1cm}
\begin{minipage}[b]{0.49\linewidth}
\centering
\includegraphics[width=8.7cm, height=6.63cm]{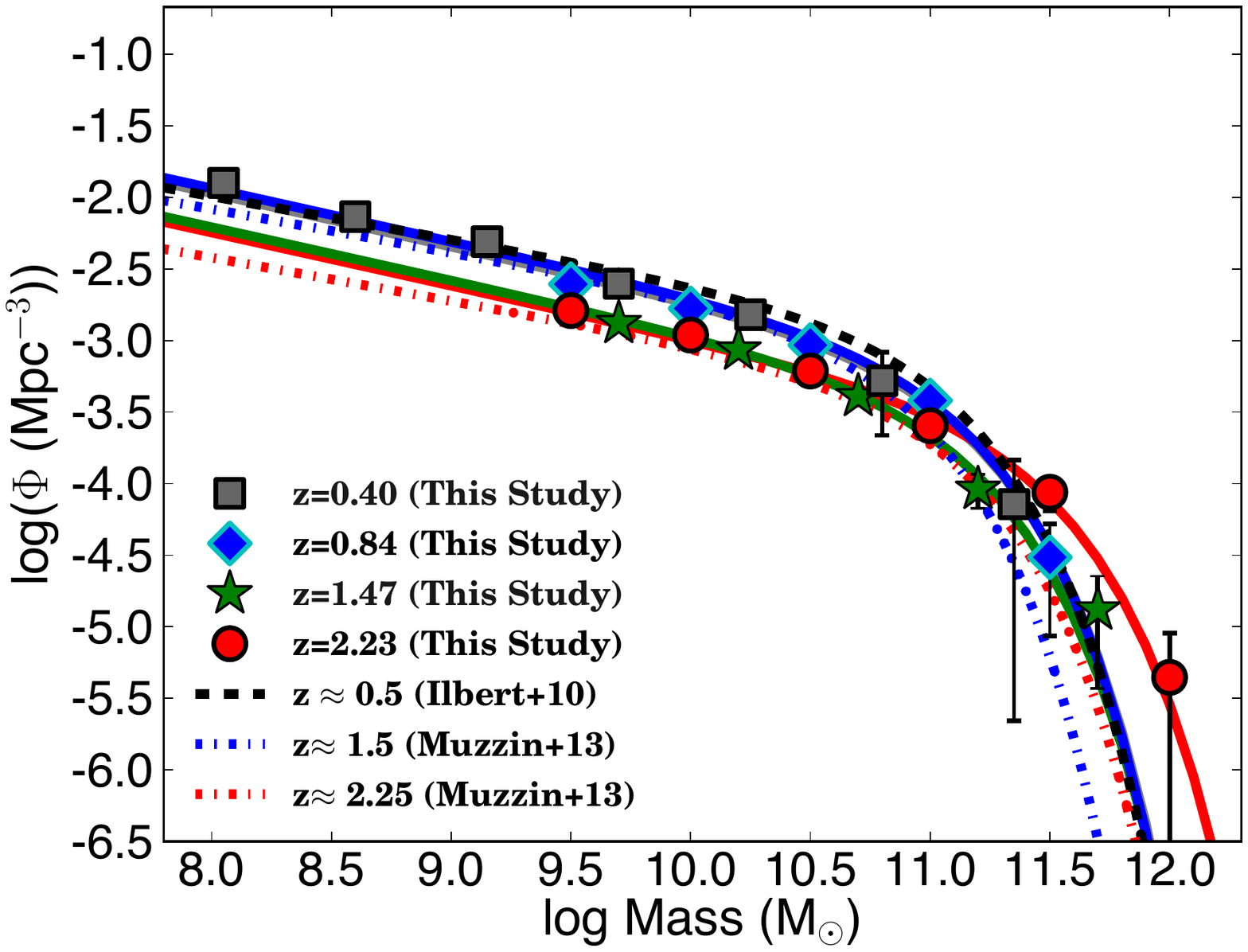}
\end{minipage}
\caption[EWs1]{{\it Left}:The star formation rate function (after extinction correction) evolution up to $z=2.23$. This reveals the strong evolution in SFR$^*$ with redshift. The results agree reasonably well with \cite{Sobral13} which assume a single dust correction, but show some differences at the bright and faint ends. This means that the more sophisticated dust correction leads to slightly higher SFR$^*$ values at all redshifts (particularly higher at $z=2.23$), but also to slightly lower $\Phi^*$ values. {\it Right}: The stellar mass function and its evolution for star-forming galaxies using our H$\alpha$ selected sample of star-forming galaxies since $z=2.23$. This shows that over the last 11 billion years the mass function of star-forming galaxies has been remarkably constant, particularly given the evolution of the star formation function since $z=2.23$. For comparison, we also show stellar mass functions (for star-forming galaxies) from the literature (selected from either observed $K$ or 3.6\,$\mu$m, while the star-forming selection is mostly a colour-colour selection), which agree well with our mass functions.   \label{SFR_mass_function} \label{SFR_function} \label{mass_function}}
\end{figure*}

\subsection{The Evolution of the SFR Function}  \label{SFR_11GYR}

The SFR function and its evolution with redshift is shown in Figure \ref{SFR_function}. We also list the best Schechter fit parameters in Table \ref{SFR_FUNCTION} (which characterise the SFR functions by their faint-end slope, $\alpha$, characteristic SFR at the break of the SFR function, SFR$^*$, and the number density of galaxies at SFR$^*$, given by $\Phi^*$).  The results reveal a strong evolution in the typical SFR of galaxies (SFR$^*$), which can be modelled very accurately up to $z=2.23$ as a function of time with the following simple relation:

\begin{equation}
   {\rm SFR}^{{\rm *}}(T[{\rm Gyr}])=10^{4.23/T+0.37} \rm M_{\odot}\,yr^{-1},
\end{equation}
It can also be given as a function of redshift by:
\begin{equation}
   {\rm SFR}^{{\rm *}}(z)=10^{0.55z+0.57} \rm M_{\odot}\,yr^{-1}.
\end{equation} 
While our parameterisation of SFR$^*$ is based on our samples at $0.4 \lsim z \lsim2.23$, it is also possible to verify if it also holds down to $z\sim0$. \cite{Bothwell} computed the SFR function at $z=0.05$, finding SFR$^*=5.0$\,M$_{\odot}$\,yr$^{-1}$, while our parameterisation of SFR$^*$($T$=12.8 Gyr) predicts 5.0\,M$_{\odot}$\,yr$^{-1}$. We therefore conclude that the evolution of SFR$^*$ is valid for $0 \lsim z \lsim2.23$. We also note that this could be interpreted as the overall star-forming population mostly being driven by a statistical typical star-formation rate (SFR$^*$) which is exponentially declining with time.

The normalisation of the SFR function, $\Phi^*$, reveals an increase from $z\sim0$ to $z\sim1$ and a further decrease afterwards, making $\Phi^*$ at $z=2.23$ roughly the same as at $\sim0.4$, but with an evolution of an order of magnitude in SFR$^*$. The evolution of $\Phi^*$ can be parameterised (with T in Gyr) by:

\begin{equation}
\log_{10}(\Phi^*[\rm T]) = 0.004231T^3-0.1122T^2+0.858T-4.659
\end{equation}

The faint-end slope is found to scatter around $\alpha\sim-1.6$. Our results therefore agree well with \cite{Smit} and show that even after a more sophisticated dust correction to the H$\alpha$ luminosity function, the faint-end slope is still consistent with being constant and the main evolution is seen in SFR$^*$.

When compared to the H$\alpha$ LF converted directly to SFR using a simple $A_{\rm H\alpha}=1$\,mag extinction correction (as in \citealt{Sobral13}; see Figure \ref{mass_function}) we find a reasonable good agreement, revealing that for the purposes of determining the bulk of the luminosity function evolution and $\rho_{\rm SFR}$, a simple uniform statistical correction is comparable to a more sophisticated one. However, at high masses the dust extinction affecting star-forming galaxies is higher, leading to a boost in the number of strongly star-forming galaxies (much closer to estimates from IR/FIR number densities; e.g. \citealt{Swinbank13}), while at the lowest masses the typical dust extinction is lower than that obtained with a uniform correction.

\begin{figure}
\centering
\includegraphics[width=8.2cm]{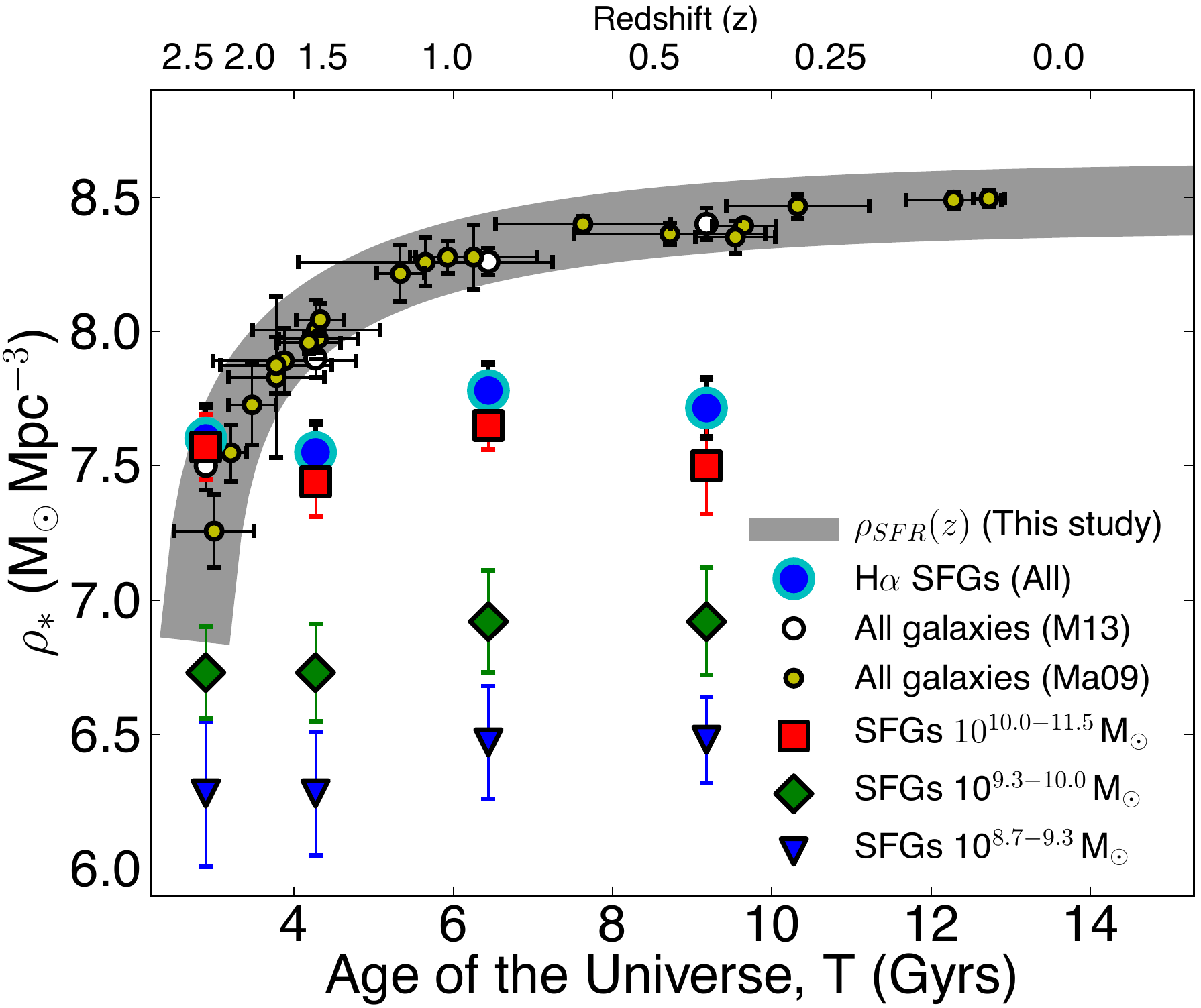}
\caption[Mass function]{The evolution of the stellar mass density for star-forming galaxies. This shows the stellar mass density for the entire sample of star-forming galaxies, revealing a roughly constant or slow rising behaviour, but also shows that the same behaviour is seen when the sample is split in different mass bins. We also show a compilation of stellar mass density determinations at various redshifts (all galaxies, regardless of star-forming or not; Marchesini et al. 2009 [Ma09]; Muzzin et al. 2013[M13]; Ilbert et al. 2013). Our prediction of the global evolution of the stellar mass density based on our SFR functions and the $\rho_{\rm SFR}$ obtained from them is shown in grey ($\rho_{SFR}(z)$), revealing an excellent agreement with observations. \label{SMD_evo}}
\end{figure}

The SFR functions derived at different redshifts allow us to evaluate $\rho_{\rm SFR}$ and its evolution. We obtain $\rho_{\rm SFR}$ by integrating our SFR functions and removing a potential contribution from AGN of 10\% at $z=0.4$ and $z=0.84$ \citep[see][]{Garn2010a} and 15\% at both $z=1.47$ and $z=2.23$, following \cite{Sobral13} -- see also \cite{Stott13b}. Similarly to what has been found by \cite{Sobral13}, the $\rho_{\rm SFR}$ is found to rise with redshift as $\log_{10}\rho_{\rm SFR}(T,\rm Gyr)=-0.136T-0.5$. This is fully consistent with the evolution of the stellar mass density (see Figure \ref{SMD_evo}), including new results from \cite{Muzzin13} and \cite{Ilbert} \footnote{Recent studies such as \cite{Ilbert} conclude that there is a significant discrepancy between the evolution of the stellar mass density and the star formation rate density, but we note that such discrepancy is only found at $z>2.2$. Thus, if we start at $z=0$ with the present day stellar mass density and remove stellar mass density of the Universe based on $\rho_{\rm SFR}$($z$), then the evolution of the stellar mass density is fully reproduced by our star formation history (see Figure \ref{SMD_evo}). The problem is that by starting at very high redshift and integrating to predict the evolution of the stellar mass density, one propagates the discrepancy at $z>2.2$ to $z<2.2$.}. 

%
%
%
\begin{table*}
 \centering
  \caption{The star formation rate (SFR) function (after correcting for extinction using stellar mass) and star formation rate density evolution for $0.4<z<2.2$ (Chabrier IMF). The measurements are obtained at $z=0.4$, $0.84$, $1.47$ and $2.23$. Columns present the redshift, break of the luminosity function, SFR$^*$ (Chabrier IMF), normalisation ($\Phi^*$) and faint-end slope ($\alpha$) of the SFR function. $\rho_{\rm SFR (>1.5)}$ and $\rho_{\rm SFR (All)}$ present the star formation rate density at each redshift based on integrating the SFR function down to $\approx1.5$\,M$_{\odot}$\,yr$^{-1}$ and for a full integration, respectively. Star formation rate densities include a correction for AGN contamination of 10\% at $z=0.4$ and $z=0.84$ \citep[see][]{Garn2010a} and 15\% at both $z=1.47$ and $z=2.23$, following Sobral et al. (2013). Errors on the faint-end slope $\alpha$ are the 1\,$\sigma$ deviation from the best fit, when fitting the three parameters simultaneously. Since the faint-end slope is more poorly constrained (specially due to the mass-dependent extinction correction), the values of SFR$^*$ and $\Phi^*$ and their errors are calculated by fixing $\alpha=-1.6$.}
  \begin{tabular}{@{}ccccccc@{}}
  \hline
   \bf Redshift & $\bf SFR^*_{\alpha=-1.6}$ & $\rm \bf \Phi^*_{\alpha=-1.6}$ & $\bf \alpha$ & $\bf \rho_{\rm \bf SFR (>1.5)}$  & $\bf \rho_{\rm \bf SFR (All)}$  \\
     ($z$)    & (M$_{\odot}$\,yr$^{-1}$) & (Mpc$^{-3}$) &    & (M$_{\odot}$\,yr$^{-1}$ Mpc$^{-3}$)   & (M$_{\odot}$\,yr$^{-1}$ Mpc$^{-3}$)  \\
 \hline
   \noalign{\smallskip}
$z= 0.40\pm0.01$ & $  6.2^{+  1.0}_{-  0.9}$ & $-3.04^{+ 0.05}_{- 0.06}$ & $-1.59^{+ 0.08}_{-0.08}$ & $0.004^{+0.0004}_{-0.0003}$  & $0.012^{+0.001}_{-0.001}$ \\
$z= 0.84\pm0.02$ & $ 10.4^{+  0.3}_{-  0.8}$ & $-2.66^{+ 0.03}_{- 0.03}$ & $-1.68^{+ 0.14}_{-0.13}$ & $0.020^{+0.001}_{-0.001}$  & $0.045^{+0.001}_{-0.001}$ \\
$z= 1.47\pm0.02$ & $ 24.5^{+  1.4}_{-  1.3}$ & $-2.71^{+ 0.05}_{- 0.05}$ & $-1.58^{+ 0.14}_{-0.29}$ & $0.055^{+0.002}_{-0.002}$  & $0.095^{+0.006}_{-0.006}$ \\
$z= 2.23\pm0.02$ & $ 65^{+  3}_{-  3}$ & $-3.05^{+ 0.03}_{- 0.03}$ & $-1.74^{+ 0.13}_{-0.14}$ & $0.083^{+0.001}_{-0.001}$  & $0.116^{+0.002}_{-0.002}$ \\
 \hline
\end{tabular}
\label{SFR_FUNCTION}
\end{table*}

%
%
\begin{table*}
 \centering
  \caption{The stellar mass function (Chabrier IMF) of H$\alpha$ selected galaxies and stellar mass density in star-forming galaxies for $0.4\sim z\sim 2.2$. The measurements are obtained at $z=0.4$, $0.84$, $1.47$ and $2.23$. Columns present the redshift, break of the stellar mass function, M$^*$, normalisation ($\Phi^*$) and faint-end slope ($\alpha$) of the stellar mass function. The two right columns present the stellar mass density at each redshift based on integrating the mass function down to $\approx10^{9.5}$\,M$_{\odot}$ and for a full integration. Errors on the faint-end slope $\alpha$ are the 1\,$\sigma$ deviation from the best fit, when fitting the three parameters simultaneously. In order to minimise the errors, and as $\alpha$ is not found to evolve significantly, $\alpha$ is fixed at the $z=0.4$ value (where the determination is most robust) and thus M$^*$ and $\Phi^*$ are obtained by fixing $\alpha=-1.37$. The last column presents the fraction of stellar mass density which is found in star-forming galaxies: this is obtained by the dividing the stellar mass density within star-forming galaxies by the total stellar mass density when using the full population \citep[from][]{Marchesini,Muzzin13}. The results show that the fraction of stellar mass density locked in star-forming galaxies is a very strong function of redshift or cosmic time.}
  \begin{tabular}{@{}cccccccc@{}}
  \hline
   \bf Redshift & $\bf M^*_{\alpha=-1.37}$ & $\rm \bf \Phi^*_{\alpha=-1.37}$ & $\bf \alpha$ & $\log_{10}$\, $\bf \rho_{\rm \bf *}$ \bf $\bf >9.5$  & $\log_{10}$\, $\bf \rho_{\rm \bf *}$ \bf All  & $\rho_*$(SFGs)/$\rho_*$ \\
     ($z$)    & $\log_{10}$\,(M$_{\odot}$) & (Mpc$^{-3}$) &    & (M$_{\odot}$\,Mpc$^{-3}$)   & (M$_{\odot}$\,Mpc$^{-3}$) & (\%)  \\ 
 \hline
   \noalign{\smallskip}
$z= 0.40\pm0.01$ & $11.07^{+ 0.54}_{- 0.54}$ & $-3.45^{+ 0.05}_{- 0.20}$ & $-1.37^{+ 0.02}_{- 0.02}$ & $7.643^{+0.046}_{- 0.04}$  & $7.715^{+0.06}_{- 0.06}$ & $19\pm3$ \\
$z= 0.84\pm0.02$ & $11.17^{+ 0.08}_{- 0.08}$ & $-3.55^{+ 0.03}_{- 0.04}$ & $-1.30^{+ 0.05}_{- 0.06}$ & $7.735^{+0.023}_{- 0.02}$  & $7.780^{+0.02}_{- 0.02}$ & $26\pm5$ \\
$z= 1.47\pm0.02$ & $11.11^{+ 0.05}_{- 0.05}$ & $-3.71^{+ 0.03}_{- 0.03}$ & $-1.37^{+ 0.06}_{- 0.06}$ & $7.496^{+0.023}_{- 0.02}$  & $7.546^{+0.02}_{- 0.02}$ & $36\pm8$ \\
$z= 2.23\pm0.02$ & $11.37^{+ 0.08}_{- 0.08}$ & $-3.82^{+ 0.03}_{- 0.04}$ & $-1.38^{+ 0.05}_{- 0.04}$ & $7.568^{+0.029}_{- 0.03}$  & $7.602^{+0.03}_{- 0.03}$ & $100^{+0}_{-20}$ \\
 \hline
\end{tabular}
\label{MF_FIT}
\end{table*}

%
%

%
%
\begin{figure*}
\centering
\includegraphics[width=17.2cm]{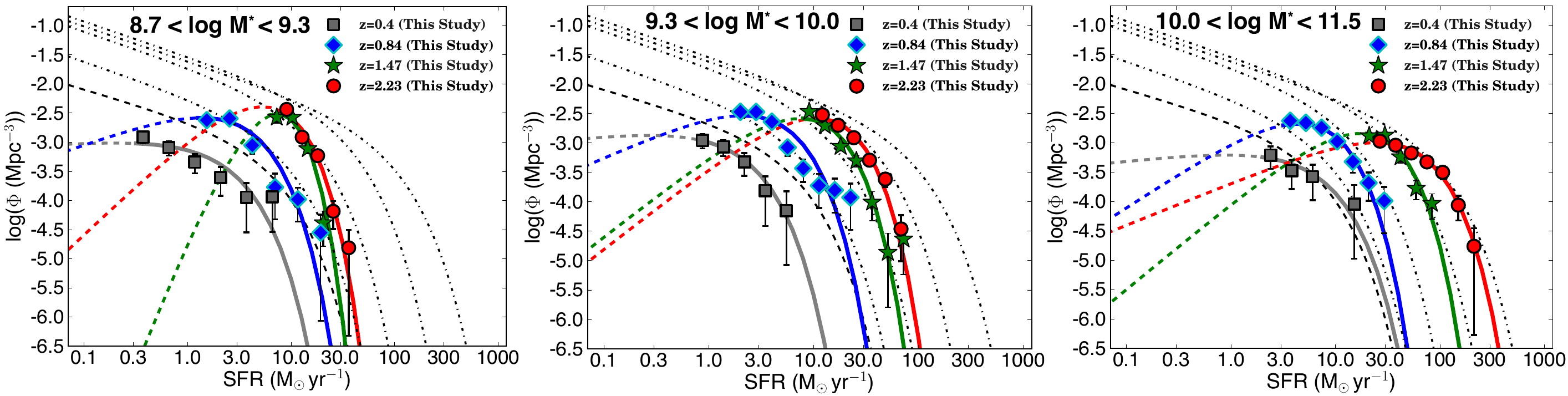}
\caption[Mass function]{The SFR function evolution for star-forming galaxies with different stellar masses. This shows that the evolution of the SFR function, mostly driven by an increase of SFR$^*$, is also seen for star-forming galaxies of different masses. Moreover, the data show that while more massive star-forming galaxies dominate at high SFRs (presenting higher SFR$^*$, but lower $\Phi^*$), lower mass star-forming galaxies dominate the SFR function at low SFRs, having a higher $\Phi^*$ and lower SFR$^*$. The results also show that the differences between the SFR function for different masses are, to first order, maintained throughout $0.4\sim z\sim2.23$, showing no significant increase in the importance of lower or higher mass galaxies in setting the SFR function. For reference, and as a comparison, the dot-dashed line presents the total SFR functions derived in this paper, while the dashed line shows the SFR function for $z\sim0$ from \cite{Bothwell}. 
\label{SFR_split_mass}}
\end{figure*}

\subsection{The Mass Function of star-forming galaxies}  \label{MASS_function}

The results, presented in Figure \ref{mass_function} show that, to first order, there is relatively little evolution in the stellar mass function of star-forming galaxies. In Table \ref{MF_FIT} we show the best-fit Schechter parameters at each redshift. There is a slight evolution in M$^*$, from $\sim10^{11}$\,M$_{\odot}$ at $z=0.40$ to $10^{11.4}$\,M$_{\odot}$ at $z=2.23$. M$^*$ is in remarkably good agreement (within the errors) with those $K$ or 3.6\,$\mu$m selected samples of non-quiescent galaxies and further strengthens the result that since $z\sim2$, M$^*$ has evolved very little (less than a factor of two), and is fixed at a few times the Milky Way mass. Thus, the bulk of the evolution in the mass function of star-forming galaxies in the last 9 Gyrs is a slight increase in the normalisation, $\Phi^*$, in agreement with e.g. \cite{Ilbert} and \cite{Muzzin13}.

The faint-end slope is found to be $\alpha=-1.37\pm0.03$ at $z=0.40$ and $\sim-1.3$ to $\sim-1.4$ at the other redshifts. For that reason, we decide to fix $\alpha$ of the mass function of star-forming galaxies to $-1.37$ to allow simpler comparisons. This is a value which is consistent with that found by other studies \citep[e.g.][]{Marchesini,Ilbert,Peng,Muzzin13}. 

We also compare our results with mass-selected mass functions of colour-colour selected star-forming galaxies at $z=0.3-0.5$ from the COSMOS survey \citep{Ilbert} and at $z\sim1-2$ from UltraVISTA \citep[][]{Muzzin13} in Figure \ref{mass_function}. The comparison between our $z=0.4$ mass function and that of Ilbert et al. reveals that our completeness corrections appear to work well, as our total stellar mass function for star-forming galaxies at $z\sim0.4$ is able to recover the mass-selected mass function. The $z\sim0.5$ mass function of \cite{Muzzin13} presents a higher normalisation, but a consistent M$^*$, and thus the difference is most likely being driven by sample (cosmic) variance. Furthermore, while there is good agreement between our results and those of Ilbert et al. and Muzzin et al. at $z\sim1-2$, we still find a slightly higher normalisation than those studies at $z\sim2.23$. The volumes at $z\sim2.2$ are relatively large, so the errors due to cosmic variance are likely to be much smaller than at lower redshift. Instead, the differences here are most likely being driven by the different selection of star-forming galaxies. Both Ilbert et al. and Muzzin et al. use the $UVJ$ selection \cite[c.f.][]{Wuyts} for flagging $K$-selected (observed) galaxies as star-forming, which has yet to be calibrated/tested to select $z\sim2$ star-forming galaxies.

\subsubsection{Stellar mass density in star-forming galaxies}  \label{SM_density_SFGs}

By integrating the mass function for star-forming galaxies, we can estimate the stellar mass density in star-forming galaxies. The results are shown in Figure \ref{SMD_evo} and in Table \ref{MF_FIT} and show that the stellar mass density in star-forming galaxies is roughly constant across 11 billion years at $\sim10^{7.65\pm0.08}$\,M$_{\odot}$\,Mpc$^{-3}$. This is different from the evolution of the stellar mass density in the Universe for galaxies as a whole (all galaxies; see Figure \ref{SMD_evo}), that evolves strongly \citep[e.g.][]{Marchesini,Ilbert,Muzzin13} from $z=2.23$ to $z=0.4$ (and that is fully reproduced by the evolution of our SFR functions -- see Figure \ref{SMD_evo}). Thus, our results imply that the fraction of stellar mass density locked up in star-forming galaxies quickly declines from virtually $\sim100$ per cent at $z\sim2.2$ to only $\sim20$ per cent at $z\sim0.4$. Our results imply that the Universe at $z=2.23$ (the likely peak of the star formation history of the Universe) was very different, with the bulk of the stellar mass density being in galaxies that were still producing stars. Our results imply a significant rise of the stellar mass density in quenched galaxies, and thus the increase of the quenched population in the last 11 billion years.

We also split the sample in three stellar mass bins ($\log_{10}$\,M\,[M$_{\odot}$]): $ 9.00\pm 0.30$, $ 9.65\pm 0.35$ and $10.75\pm 0.75$ and find that there is also relatively little evolution in the stellar mass density of star-forming galaxies with different masses (Figure \ref{SMD_evo}).

\subsection{SFR functions and contribution to $\bf \rho_{\rm SFR}$ from different masses: downsizing?}  \label{rho_SFR_diff_MASSES}

\begin{figure*}
\begin{minipage}[b]{0.49\linewidth}
\centering
\includegraphics[width=8.7cm,height=7.1cm]{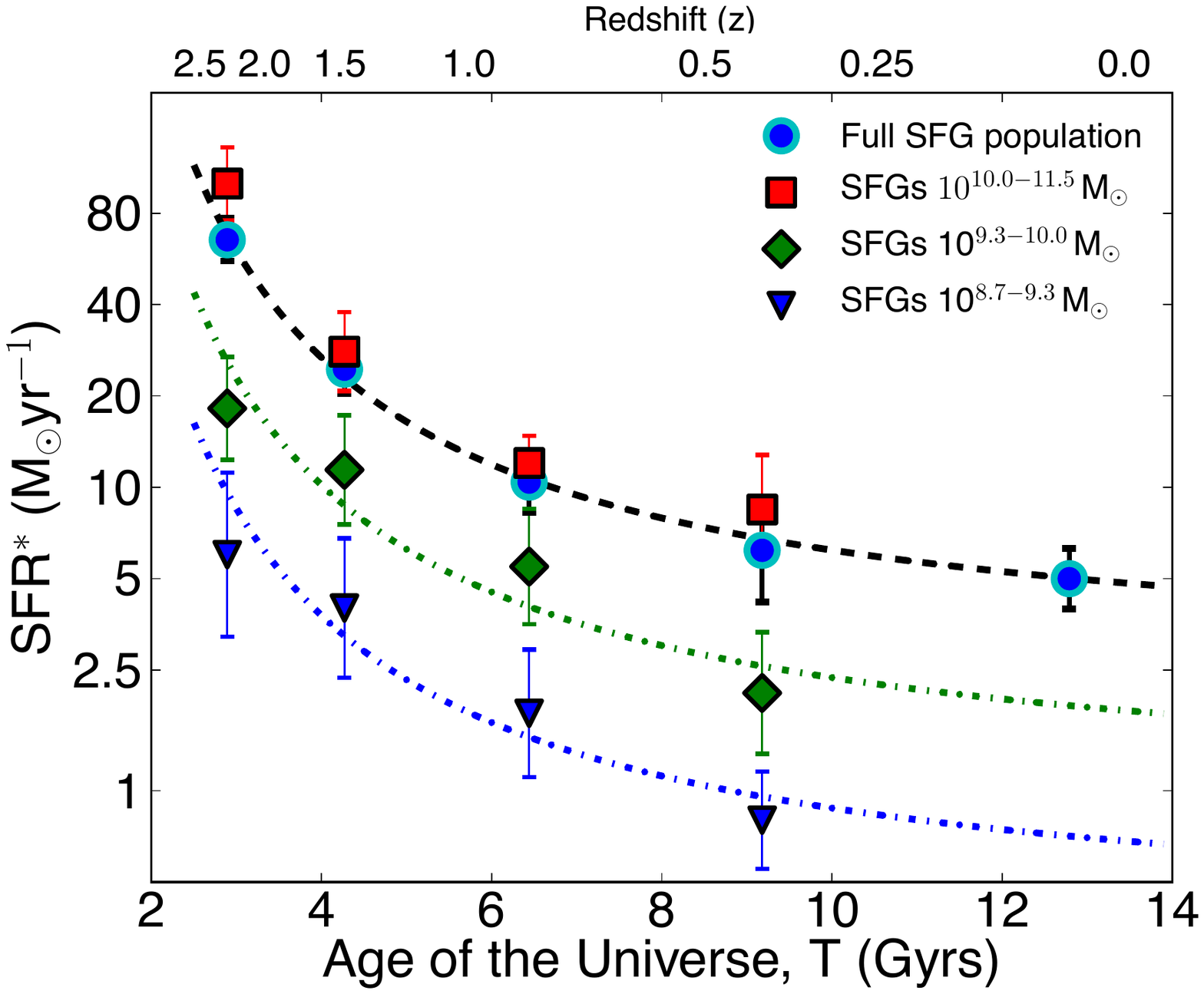}
\end{minipage}
\hspace{0.1cm}
\begin{minipage}[b]{0.49\linewidth}
\centering
\includegraphics[width=8.7cm,height=7.1cm]{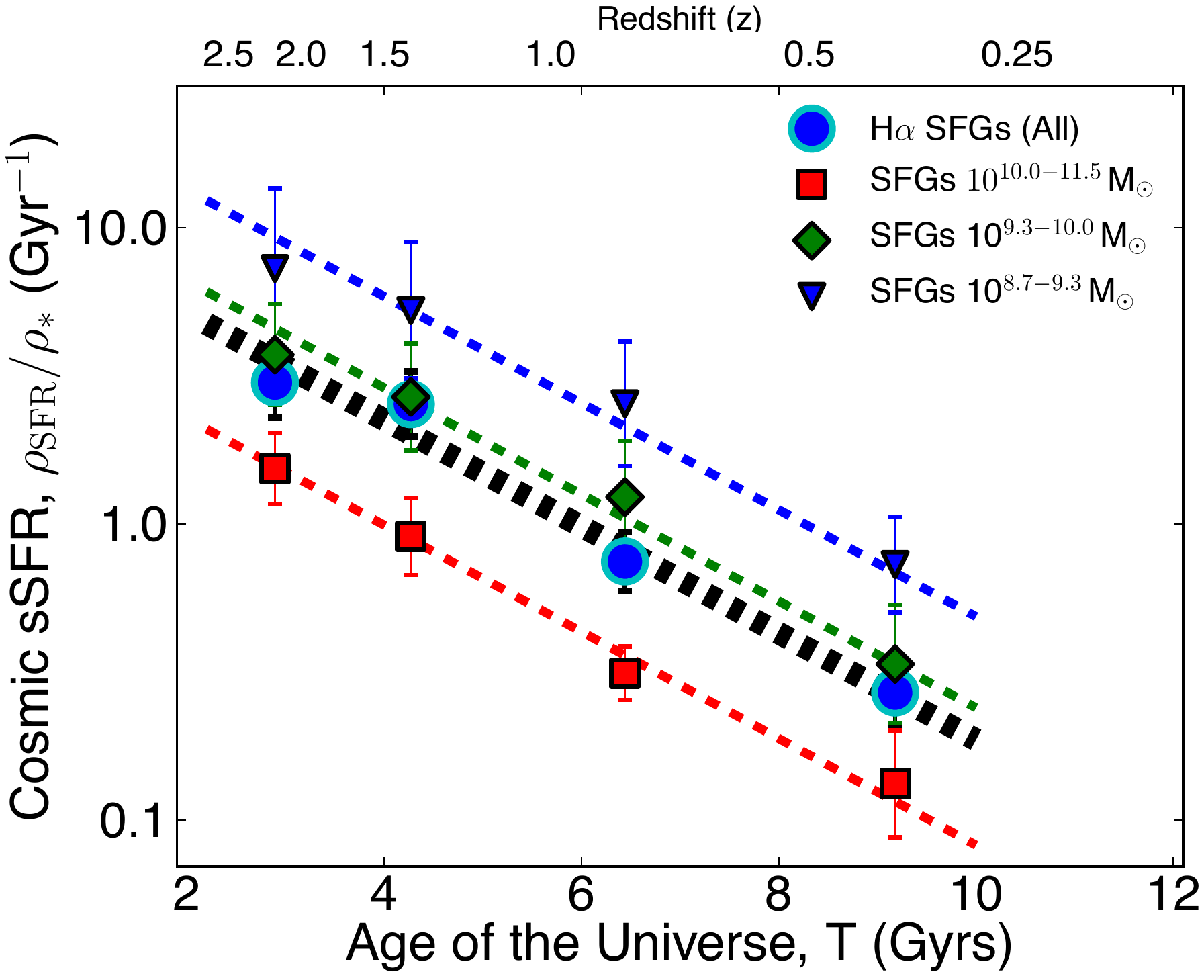}
\end{minipage}
\caption[EWs1]{{\it Left}: The characteristic SFR, SFR$^*$, as a function of time/redshift, for the entire population of star-forming galaxies and for sub-samples with different masses. The fit to the full star-forming population is given by Equation 3. This functional form is also found to provide a very good fit to the sub-samples with a change in its normalisation. This shows that the reduction of the typical SFR is occurring in a broadly self-similar way at all masses probed, and that the decrease of SFR$^*$ with time is probably driven by the decline of SFR$^*$ at all masses. The data-point at $z\sim0$ is from \cite{Bothwell}. {\it Right:} The evolution of cosmic sSFR, the ratio between the star formation rate density, $\rho_{\rm SFR}$ and the stellar mass density, $\rho_*$, as a function of cosmic time, for different stellar mass sub-samples. This shows that the cosmic specific star formation rate declines with time for star-forming galaxies at all masses. The decline happens at roughly the same rate with time at all masses, as the best fitted slopes are all $<1\sigma$ away from each other, and all consistent with that of the full sample, sSFR[$T$, Gyr]\,$\propto$\,10$^{(-0.18\pm0.03)T}$. 
 \label{SFR_mass_function} \label{COSMIC_SSFR_evo}}
\end{figure*}

We present SFR functions split by stellar mass in Figure \ref{SFR_split_mass} (see also Table \ref{SFRD_diff_MASSES} which presents the best Schechter fits to our SFR functions). We find that star-forming galaxies with higher masses present, on average, both a higher SFR$^*$ (see Figure \ref{SFR_split_mass} and Figure \ref{COSMIC_SSFR_evo}) and a lower $\Phi^*$. The SFR$^*$ of the highest mass sample is in very good agreement with that estimated for the full SFR function at each redshift, showing that these are the star-forming galaxies that are responsible for setting SFR$^*$ at each epoch. However, for any fixed stellar mass bin, we find that SFR$^*$ declines with redshift, showing that the decline of the typical SFR is not just happening for massive star-forming galaxies, but rather for all masses probed (see Figure \ref{COSMIC_SSFR_evo}). We note that this is recovered both when $\alpha$ is allowed to vary, but also find this to be the case for any fixed value of $\alpha$ from $1.0$ to $-1.6$. SFR$^*$ values increase for steeper $\alpha$ and decrease for shallower $\alpha$, but they vary consistently for all masses, and thus the differences between masses are maintained, within the errors.

%
%

For any given mass bin, the decline of SFR$^*$ with redshift seems to be relatively self-similar (see Figure \ref{COSMIC_SSFR_evo}). In order to test this we fit the decline of SFR$^*$ with redshift for the different stellar mass sub-samples with the same functional form as the SFR$^*$ decline for the full population (Equation 3), allowing for the normalisation to change. We find this provides a very good fit to all sub-samples (see Figure \ref{COSMIC_SSFR_evo}), with a reduced $\chi^2\lsim1$, confirming that the evolution is consistent with being mostly self-similar at all masses. In order to further test any difference in the SFR$^*$ evolution for different masses, we also allow the time-dependence to vary. The fits with two parameters are only marginally better than those with one parameter, providing only weak evidence of a quicker decline of SFR$^*$ for the most massive star-forming galaxies than for the lowest mass galaxies. The statistical significance of this is very weak (at the $\sim1\sigma$ level), and almost entirely driven by the $z=2.23$ sample. If the $z=2.23$ sample is removed from the analysis the weak statistical hint completely disappears, although we note that if the $z=0.4$ sample is removed instead the statistical significance increases slightly, but is still relatively weak ($\sim1.5$\,$\sigma$). We therefore conclude that there may be small differences in the decline of the typical SFR (SFR$^*$) of star-forming galaxies with different masses across redshift (particularly at $z>2$), but that the bulk of the evolution happens at the same rate at all masses probed. Our results also suggests a statistical relation between the stellar mass of star-forming galaxies and their typical SFR. For this study, we find SFR$^*\propto$\,M$^{0.56\pm0.05}$, with the exponent being relatively independent of redshift at least for $z=0.4-2.23$ (but the normalisation declines with cosmic time) and resulting in a slope ($\beta$) between $\log_{10}$sSFR and $\log_{10}$M of $\sim-0.44\pm0.05$.
 
\begin{figure}
\centering
\includegraphics[width=8.0cm]{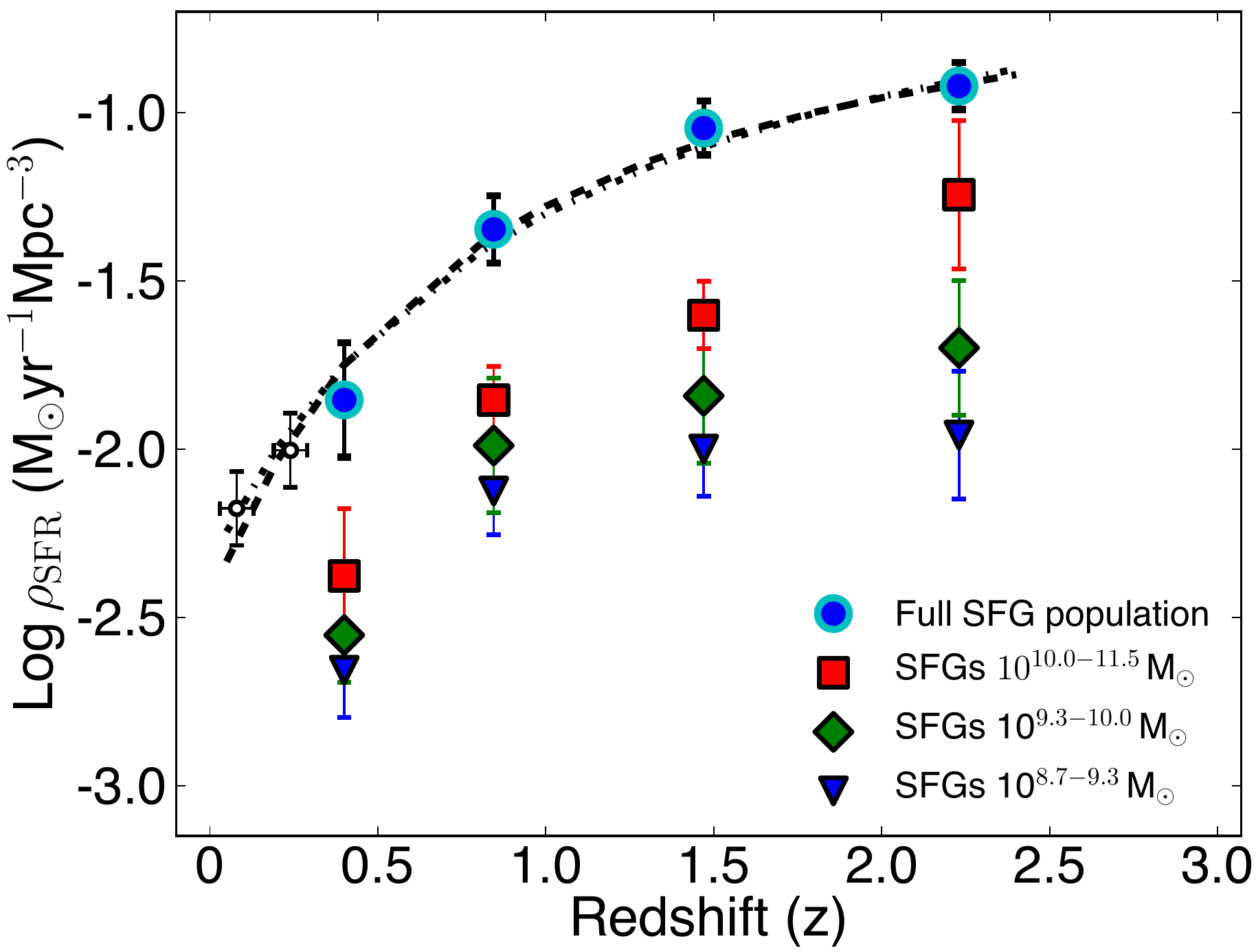}
\caption[EWs1]{The star-formation history of the Universe from this study and the contribution from star-forming galaxies with different stellar. Measurements at $z=0.07$ and $z=0.24$ are from \cite{Shioya} and \cite{Ly2007}. The results show that the contributions from galaxies with different masses do not change significantly across time, and that the decline in SFRD is seen at all masses in a self-similar way, within the errors. This is consistent with the decline in SFR* with time also seen for all three mass bins probed in this study. 
 \label{SFR_mass_function} \label{COSMIC_SFRHistory}}
\end{figure}

Even more interesting is the possibility of splitting the total $\rho_{\rm SFR}$ in the contributions from star-forming galaxies with different masses by using the SFR functions derived here. We show the results in Figure \ref{COSMIC_SFRHistory}. Our results reveal that the contributions from sub-samples with different masses are relatively constant over time/redshift. Indeed, there is a decrease of $\rho_{\rm SFR}$ at all masses, with 10$^{10-11.5}$\,M$_{\odot}$ star-forming galaxies being the bulk contributors to the total $\rho_{\rm SFR}$ within the observed population. We note that there is a slight increase of the fractional contribution to $\rho_{\rm SFR}$ from 10$^{10-11.5}$\,M$_{\odot}$ SFGs beyond $z\sim2$ (see Figure 8), but that this is seen at the $\approx$\,1$\sigma$ level only. Star-forming galaxies with masses $>10^{8.7}$\,M$_{\odot}$ account for $\sim60-70$\% of the total $\rho_{\rm SFR}$, while (due to $\alpha=-1.6$ for the SFR function) $<10^{8.7}$\,M$_{\odot}$ star-forming galaxies are expected to contribute the remaining 30\%, with this fraction being relatively constant (within the errors) at all redshifts probed.

%
%

\begin{figure}
\centering
\includegraphics[width=8.0cm]{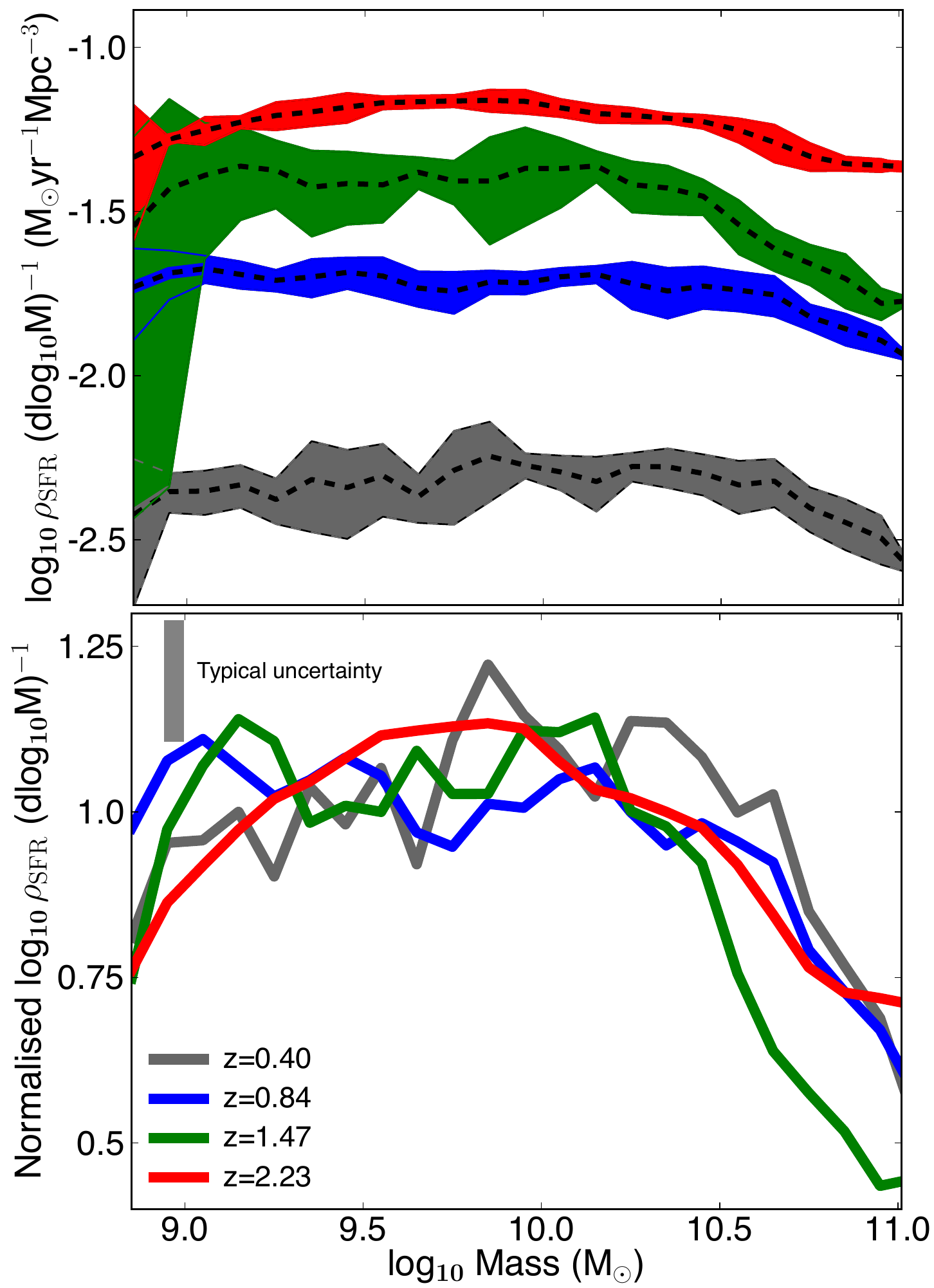}
\caption[EWs1]{{\it Top}: The star formation rate density $\rho_{\rm SFR}$ per d$\log_{10}$M as a function of mass for the four different redshifts. The results show that there is an overall increase at all masses with redshift, revealing little to no evolution in the fractional contribution from each mass. {\it Bottom}: The normalised distribution of $\rho_{\rm SFR}$ per d$\log_{10}$M (after accounting for the overall redshift evolution to allow for a clearer comparison, i.e. after dividing by the median at each redshift). We find that the normalised distribution is very similar across redshifts, revealing a peak at around 10$^{10}$\,M$_{\odot}$, and declining at the highest and lowest masses. We see no strong evolution in the shape of this function with redshift, as the small differences are all within the errors.
 \label{SFR_mass_function} \label{CONT_PER_DLOGM}}
\end{figure}

To further investigate the contributions from star-forming galaxies with different masses to the $\rho_{\rm SFR}$ and any potential evolution, and in order to obtain measurements which are independent of the bin choice (centre and width) in stellar mass, we also derive the contribution per d$\log_{10}$\,M at different masses to $\rho_{\rm SFR}$. We do this by obtaining 200 realisations of the SFR functions per redshift, following the method we have described before (fits constrained by number density of star-forming galaxies within that mass range from our stellar mass functions). Each of our realisations is obtained with a different mass bin, with the centre of the bin ranging from 10$^{8.7}$\,M$_{\odot}$ to 10$^{11.0}$\,M$_{\odot}$ and with the width of the bin ($\Delta\log_{10}$\,M) varying between of 0.3 to 1.0. We note that because of the very small number of star-forming galaxies with masses $>$10$^{11.0}$\,M$_{\odot}$ (see Figure 1) we do not allow the bin centre to be higher than 10$^{11.0}$\,M$_{\odot}$, as only very wide bins would allow for the necessary statistics, and they are dominated by 10$^{11.0}$\,M$_{\odot}$ and lower mass galaxies. For each realisation we obtain the integral of the SFR function per $\Delta\log_{10}$\,M. The results, shown in Figure \ref{CONT_PER_DLOGM}, show that $\rho_{\rm SFR}$ per d$\log_{10}$\,M increases with redshift at all masses, confirming our results when choosing 3 specific mass bins. Figure \ref{CONT_PER_DLOGM} also shows that there is a tentative peak at stellar mass of about 10$^{10\pm0.25}$\,M$_{\odot}$ (after normalising the distributions at each redshift by their median), which happens at all redshifts, and a strong decline in $\rho_{\rm SFR}$(d$\log_{10}$\,M)$^{-1}$ at the highest masses at all redshifts.

%
%
%

We therefore find no strong ``downsizing" (with redshift) in terms of the relative contributions from different masses to the star formation rate density. We note that at $z=2.23$, and for the highest masses, $\rho_{\rm SFR}$ per d$\log_{10}$\,M may be decreasing with a slightly shallower gradient than at other redshifts, consistent with the weak ($\sim$\,1\,$\sigma$) trend seen in Figure 8. However, we find a strong, general decline at all masses, with the fraction contributions of star-forming galaxies with different masses remaining relatively unchanged. We certainly do not find any evidence of a strong general decrease in the contribution of higher mass star-forming galaxies, but we also note that this is a result based on star-forming galaxies down to a SFR and EW limit, not mass-selected galaxies. Our results imply that the decline of $\rho_{\rm SFR}$ as a whole must be primarily due to processes that lead to declining SFRs with time at all masses, and not just processes that happen at specific stellar masses. This means that while mass may be empirically linked with the probability of a galaxy becoming quenched, it is not the main driver of the decline of $\rho_{\rm SFR}$. Nevertheless, the probability of a galaxy becoming/being quenched depends strongly on mass \citep[e.g.][]{Peng,Sobral11}, and thus galaxy evolution seems to be driven by both a general decline of $\rho_{\rm SFR}$ at all masses (for star-forming galaxies), and processes that, in addition, are able to quench galaxies in a mass-dependent way.

Finally, we can use our sub-samples with different masses and derive cosmic specific star-formation rates, to further test whether they are also declining in a self-consistent way. We do this very simply by dividing $\rho_{\rm SFR}$ for a given redshift and mass sub-sample by their $\rho_*$, i.e., the integral of the stellar mass function within the mass limits of the sample. The results are shown in Figure \ref{COSMIC_SSFR_evo}, revealing a general decline of the cosmic sSFR for all masses and for the full population. The decline of the cosmic sSFR (full population) with decreasing redshift is well parameterised as $(1+z)^{-3.5\pm0.5}$, but such decline is also very well parameterised as a function of time with a function of the form:

\begin{equation}
{\rm sSFR}[T,{\rm Gyr}]\,=\,10^{A\times T+B} {\rm Gyr^{-1}}.
\end{equation}
We find that the best fit is given by $A=-0.18\pm0.02$ and $B=1.07\pm0.15$ (see Figure \ref{COSMIC_SSFR_evo}). We fit the sub-samples with the same function and allow for both parameters to vary in order to investigate any significant difference between the sub-samples. We find that the best fit $A$ varies between 0.17 and 0.18 for the three sub-samples (with $\Delta A\sim0.03$ per fit), and thus all fits are well within 1$\sigma$ of each other, and all fully consistent with the value found for the full sample (see Figure \ref{COSMIC_SSFR_evo}). We also repeat the fits by excluding either the $z=2.23$ or the $z=0.4$ samples. We find that excluding the $z=2.23$ makes the best fit $A$ of the different sub-samples converge even more to $A\approx0.18$, while excluding the $z=0.4$ sample leads to a slightly wider range of $A$, with the most massive galaxies showing a steeper $A$, while the other sub-samples show a slower evolution. However, even if the $z=0.4$ sample is excluded, all fits are still within $<1$\,$\sigma$ of $A=0.18$. We therefore conclude that the cosmic sSFR is declining in a very similar way at all masses. The normalisation ($B$), however, is found to evolve very significantly from 0.7 to 1.5 from high to low masses (see Figure \ref{COSMIC_SSFR_evo}).

\section{Conclusions}  \label{conclusions}

We presented the joint evolution of the SFR and stellar mass function for star-forming galaxies since $z=2.23$, and the contribution of different masses to the star-formation history of the Universe. Our main results are:

\begin{itemize}

\item There is a significant evolution in the median EW$_0$ of H$\alpha$ emitters over the last 11 billion years. EW$_{\rm H\alpha+[NII]}$ increases as $(1+z)^{1.7\pm0.1}$ at least up to $z=2.23$, in very good agreement with other studies at lower redshift. 

\item  The SFR function is derived and shown to evolve strongly as a function of redshift/time. The bulk of the evolution is driven by the continuous increase of SFR$^*$ from $z=0$ to $z=2.23$, given by  ${\rm SFR}^{{\rm *}}(T,\rm Gyr)=10^{4.23/T+0.37}$\,M$_{\odot}$\,yr$^{-1}$. There is also a milder evolution of $\Phi^*$, increasing up to $z\sim1$, and dropping again to $z\sim2$. The faint-end slope of the SFR function, $\alpha$, is consistent with $-1.6$ and no significant evolution.

\item We derive the stellar mass function of star-forming galaxies. We show that the stellar mass function presents very little evolution over 11 Gyrs, with no evolution in the faint-end slope ($\alpha=-1.37$), revealing roughly the same M$^*\sim10^{11}$\,M$_{\odot}$ and only a mild change in the normalisation (which increases for decreasing redshift). At $z=2.23$ the stellar mass function of star-forming galaxies is characterised by a slightly higher M$^*$, consistent with $z>2$ being an epoch where even the most massive galaxies were still forming stars.

\item We find that the amount of stellar mass density in star-forming galaxies is (to first-approximation) relatively constant over time at $\sim10^{7.65\pm0.08}$\,M$_{\odot}$\,Mpc$^{-3}$. The fraction of stellar mass density contained in star-forming galaxies has been continuously declining from $\sim100$\% at $z\sim2.2$ to only $\sim20$\% at $z\sim0.4$, a consequence of the build-up of the passive/quenched population over time.

\item  $M=10^{10.0\pm0.25}$\, M$_{\odot}$ galaxies have the highest $\rho_{\rm SFR}$ per d$\log_{10}$\,M at all redshifts. Although there are weak ($\approx$\,1\,$\sigma$) indications that the relative contribution of the most massive galaxies to $\rho_{\rm SFR}$ may begin to increase beyond $z\sim2$, we find no significant evolution in the fractional contribution from SFGs of different masses to $\rho_{\rm SFR}$ or to $\rho_{\rm SFR}$(d$\log$\,M)$^{-1}$ since $z = 2.23$. Therefore, there is no significant shifting of star-formation from higher to low mass galaxies, with the results showing that the star formation activity of galaxies declines at all masses with redshift. This is seen in the decline of SFR$^*$ for all mass bins, a decline in $\rho_{\rm SFR}$(d$\log$\,M)$^{-1}$ at all masses with redshift, but also in the decline of the cosmic sSFR with time, or rise with redshift, parameterised by $(1+z)^{-3.5\pm0.5}$, up to $z\sim2.23$, which is also self-similar at all masses. These results have important implications to the main driver(s) of the declining $\rho_{\rm SFR}$ for star-forming galaxies, as such mechanism(s) must affect all masses, and not just the most massive/or least massive.

\end{itemize}

Our results point towards a simple scenario where star-forming galaxies since $z\sim2.23$ are mostly described by a continuous evolution of their typical SFR, SFR$^*$, from $\sim66$\,M$_{\odot}$\,yr$^{-1}$ to 5\,M$_{\odot}$\,yr$^{-1}$ (a factor $\sim13$ decline in 11 billion years, and a factor 10 from $z\sim2.23$ to $z\sim0.4$), while M$^*$ is kept relatively constant (relatively weak evolution) at $\sim10^{11}$\,M$_{\odot}$. It also shows that the total stellar mass density in star-forming galaxies has been roughly constant over this period, and thus the fraction of the stellar mass density in star-forming galaxies has been declining at least since $z\sim2.2$. This obviously implies that the passive/quenched galaxies and the stellar mass density contained in such galaxies have been rising significantly since $z=2.23$, in excellent agreement with e.g. \cite{Ilbert} and \cite{Muzzin13}.

The decline of the star formation activity happens at all masses, as a function of time/redshift and thus must be driven by a process that is happening at all masses and affecting the entire star-forming population, not just star-forming galaxies with a certain mass. The fact that the mass function of star-forming galaxies remains approximately constant also sheds light into the processes that may be driving galaxy evolution. Such processes need to quench both galaxies at the massive end (to keep M$^*$ constant), but also to reduce the growth of galaxies at all masses in a now very well-constrained way.

Our results can also be interpreted in the context of the results from \cite{SOBRAL10A}. They find that SFR$^*$ star-forming galaxies at redshifts up to at least $z\sim2.23$ reside in dark matter haloes with masses of $\sim10^{12}$\,M$_{\odot}$, similar to the mass of the Milky Way dark matter halo. Those results suggest that, over the last 11 Gyrs, star-forming galaxies have been hosted by (different) dark matter haloes of roughly the same masses, but that the same dark matter halo mass is only able to drive a maximum SFR which declines with redshift. This agrees well with the general results in the literature  \citep[e.g.][]{Hopkins09,Leitner12,Behroozi,Moster13}, both in the masses of the dark matter haloes that maximise galaxy formation and evolution, but also in the lack of evolution in such masses since $z\sim2$. These massive haloes are also in the transition between two modes as modelled by \cite{Oppenheimer10}. The trends often called as ``downsizing" are therefore easily interpreted in this context. Dark matter haloes similar in mass to that of the present-day Milky Way ($\sim$10$^{12}$\,M$_{\odot}$) have been the most important/efficient hosts of SFR$^*$ star-forming galaxies \citep[][]{SOBRAL10A,Geach12,Behroozi,Moster13}, but those hosting them at $z\sim2.2$ (which sustained some of the highest SFRs in the Universe, e.g. \citealt{Hickox}), have grown significantly ever since, as they had 11 Gyrs to do so, and thus, today, they are much more massive haloes, likely to host passive galaxies instead. Thus, it is not surprising that the more massive haloes found today are not star-forming, as in order to be some of the most massive haloes they had to host star-forming galaxies very early on. 

In conclusion, our results are consistent with the evolution of the galaxy population as a whole being driven by i) a general decline of star-formation rates at all masses (for star-forming galaxies), resulting in the decline of $\rho_{\rm SFR}$ as a function of cosmic time, and ii) the quenching of star-forming galaxies in a mass-dependent way, resulting in the very little evolution in the mass function of star-forming galaxies since $z=2.23$ and a strong increase in the normalisation of the mass function of quenched galaxies.

\section*{Acknowledgments}

{We thank the anonymous referee for many comments and suggestions which improved both the quality and clarity of this work. DS acknowledges financial support from the Netherlands Organisation for Scientific research (NWO) through a Veni fellowship. PNB acknowledges support from STFC. IRS and JPS thank the U.K. Science and Technology Facility Council (STFC, ST/I001573/I). IRS acknowledges the ERC Advanced Investigator programme DUSTYGAL and a Royal Society/Wolfson Merit Award. The authors wish to thank Valentino Gonzalez, Renske Smit, Adam Muzzin, Mattia Fumagalli, Shannon Patel, Fei Li and Richard Bower for fruitful discussions. The data on which this analysis is based are available from \cite{Sobral13}.

\bibliographystyle{mn2e}
\bibliography{bibliography}

\begin{thebibliography}{}

\bibitem[\protect\citeauthoryear{{Alavi}, {Siana}, {Richard}, {Stark},
  {Scarlata}, {Teplitz}, {Freeman}, {Dominguez}, {Rafelski}, {Robertson} \&
  {Kewley}}{{Alavi} et~al.}{2013}]{Alavi13}
{Alavi} A.,  {Siana} B.,  {Richard} J.,  {Stark} D.~P.,  {Scarlata} C.,
  {Teplitz} H.~I.,  {Freeman} W.~R.,  {Dominguez} A.,  {Rafelski} M.,
  {Robertson} B.,    {Kewley} L.,  2013, ApJ, submitted, arXiv:1305.2413

\bibitem[\protect\citeauthoryear{{Behroozi}, {Wechsler} \& {Conroy}}{{Behroozi}
  et~al.}{2013}]{Behroozi}
{Behroozi} P.~S.,  {Wechsler} R.~H.,    {Conroy} C.,  2013, ApJL, 762, L31

\bibitem[\protect\citeauthoryear{{Best}, {Smail}, {Sobral}, {Geach}, {Garn},
  {Ivison}, {Kurk}, {Dalton}, {Cirasuolo} \& {Casali}}{{Best}
  et~al.}{2010}]{Best2010}
{Best} P.,  {Smail} I.,  {Sobral} D.,  {Geach} J.,  {Garn} T.,  {Ivison} R.,
  {Kurk} J.,  {Dalton} G.,  {Cirasuolo} M.,    {Casali} M.,  2010, UKIRT30
  proceedings, arXiv:1003.5183

\bibitem[\protect\citeauthoryear{{Bothwell}, {Kenicutt}, {Johnson}, {Wu},
  {Lee}, {Dale}, {Engelbracht}, {Calzetti} \& {Skillman}}{{Bothwell}
  et~al.}{2011}]{Bothwell}
{Bothwell} M.~S.,  {Kenicutt} R.~C.,  {Johnson} B.~D.,  {Wu} Y.,  {Lee} J.~C.,
  {Dale} D.,  {Engelbracht} C.,  {Calzetti} D.,    {Skillman} E.,  2011, MNRAS,
  415, 1815

\bibitem[\protect\citeauthoryear{{Brammer}, {Whitaker}, {van Dokkum},
  {Marchesini}, {Labb{\'e}}, {Franx}, {Kriek}, {Quadri}, {Illingworth}, {Lee},
  {Muzzin} \& {Rudnick}}{{Brammer} et~al.}{2009}]{Brammer}
{Brammer} G.~B.,  {Whitaker} K.~E.,  {van Dokkum} P.~G.,  {Marchesini} D.,
  {Labb{\'e}} I.,  {Franx} M.,  {Kriek} M.,  {Quadri} R.~F.,  {Illingworth} G.,
   {Lee} K.-S.,  {Muzzin} A.,    {Rudnick} G.,  2009, ApJL, 706, L173

\bibitem[\protect\citeauthoryear{{Bruzual}}{{Bruzual}}{2007}]{B07}
{Bruzual} G.,  2007, in {Vallenari} A.,  {Tantalo} R.,  {Portinari} L.,
  {Moretti} A.,  eds, From Stars to Galaxies: Building the Pieces to Build Up
  the Universe Vol.~374 of Astronomical Society of the Pacific Conference
  Series, {Stellar Populations: High Spectral Resolution Libraries. Improved
  TP-AGB Treatment}.
p.~303

\bibitem[\protect\citeauthoryear{{Bruzual} \& {Charlot}}{{Bruzual} \&
  {Charlot}}{2003}]{BC03}
{Bruzual} G.,  {Charlot} S.,  2003, MNRAS, 344, 1000

\bibitem[\protect\citeauthoryear{{Calzetti}, {Armus}, {Bohlin}, {Kinney},
  {Koornneef} \& {Storchi-Bergmann}}{{Calzetti} et~al.}{2000}]{Calzetti}
{Calzetti} D.,  {Armus} L.,  {Bohlin} R.~C.,  {Kinney} A.~L.,  {Koornneef} J.,
    {Storchi-Bergmann} T.,  2000, ApJ, 533, 682

\bibitem[\protect\citeauthoryear{{Capak}, {Aussel}, {Ajiki}, {McCracken},
  {Mobasher}, {Scoville} \& {et al.}}{{Capak} et~al.}{2007}]{Capak}
{Capak} P.,  {Aussel} H.,  {Ajiki} M.,  {McCracken} H.~J.,  {Mobasher} B.,
  {Scoville} N.,    {et al.} 2007, ApJS, 172, 99

\bibitem[\protect\citeauthoryear{{Casali}, {Adamson}, {Alves de Oliveira},
  {Almaini}, {Burch}, {Chuter}, {Elliot} \& {et al.}}{{Casali}
  et~al.}{2007}]{Casali}
{Casali} M.,  {Adamson} A.,  {Alves de Oliveira} C.,  {Almaini} O.,  {Burch}
  K.,  {Chuter} T.,  {Elliot} J.,    {et al.} 2007, A\&A, 467, 777

\bibitem[\protect\citeauthoryear{{Chabrier}}{{Chabrier}}{2003}]{Chabrier}
{Chabrier} G.,  2003, PASP, 115, 763

\bibitem[\protect\citeauthoryear{{Cimatti}, {Daddi} \& {Renzini}}{{Cimatti}
  et~al.}{2006}]{Cimatti06}
{Cimatti} A.,  {Daddi} E.,    {Renzini} A.,  2006, A\&A, 453, L29

\bibitem[\protect\citeauthoryear{{Cirasuolo}, {McLure}, {Dunlop}, {Almaini},
  {Foucaud} \& {Simpson}}{{Cirasuolo} et~al.}{2010}]{Cirasuolo10}
{Cirasuolo} M.,  {McLure} R.~J.,  {Dunlop} J.~S.,  {Almaini} O.,  {Foucaud} S.,
     {Simpson} C.,  2010, MNRAS, 401, 1166

\bibitem[\protect\citeauthoryear{{Colbert}, {Teplitz}, {Atek}, {Bunker},
  {Rafelski}, {Ross} \& {et al.}}{{Colbert} et~al.}{2013}]{Colbert}
{Colbert} J.~W.,  {Teplitz} H.,  {Atek} H.,  {Bunker} A.,  {Rafelski} M.,
  {Ross} N.,    {et al.} 2013, arXiv:1305.1399

\bibitem[\protect\citeauthoryear{{Cowie}, {Songaila}, {Hu} \& {Cohen}}{{Cowie}
  et~al.}{1996}]{Cowie}
{Cowie} L.~L.,  {Songaila} A.,  {Hu} E.~M.,    {Cohen} J.~G.,  1996, AJ, 112,
  839

\bibitem[\protect\citeauthoryear{{Cucciati}, {Tresse}, {Ilbert}, {Le
  F{\`e}vre}, {Garilli}, {Le Brun}, {Cassata} \& {et al.}}{{Cucciati}
  et~al.}{2012}]{Cucciati}
{Cucciati} O.,  {Tresse} L.,  {Ilbert} O.,  {Le F{\`e}vre} O.,  {Garilli} B.,
  {Le Brun} V.,  {Cassata} P.,    {et al.} 2012, A\&A, 539, A31

\bibitem[\protect\citeauthoryear{{Damen}, {F{\"o}rster Schreiber}, {Franx},
  {Labb{\'e}}, {Toft}, {van Dokkum} \& {Wuyts}}{{Damen} et~al.}{2009}]{Damen}
{Damen} M.,  {F{\"o}rster Schreiber} N.~M.,  {Franx} M.,  {Labb{\'e}} I.,
  {Toft} S.,  {van Dokkum} P.~G.,    {Wuyts} S.,  2009, ApJL, 705, 617

\bibitem[\protect\citeauthoryear{{Dom{\'{\i}}nguez}, {Siana}, {Henry},
  {Scarlata}, {Bedregal}, {Malkan}, {Atek} \& {et al.}}{{Dom{\'{\i}}nguez}
  et~al.}{2013}]{Dominguez}
{Dom{\'{\i}}nguez} A.,  {Siana} B.,  {Henry} A.~L.,  {Scarlata} C.,  {Bedregal}
  A.~G.,  {Malkan} M.,  {Atek} H.,    {et al.} 2013, ApJ, 763, 145

\bibitem[\protect\citeauthoryear{{Dunlop}}{{Dunlop}}{2012}]{Dunlop}
{Dunlop} J.~S.,  2012, ASSSL, 396, 223, arXiv:1205.1543

\bibitem[\protect\citeauthoryear{{Ellis}}{{Ellis}}{2008}]{ELLIS}
{Ellis} R.~S.,  2008, {Observations of the High Redshift Universe}.
pp 259--364

\bibitem[\protect\citeauthoryear{{Fontanot}, {De Lucia}, {Monaco}, {Somerville}
  \& {Santini}}{{Fontanot} et~al.}{2009}]{Fontanot}
{Fontanot} F.,  {De Lucia} G.,  {Monaco} P.,  {Somerville} R.~S.,    {Santini}
  P.,  2009, MNRAS, 397, 1776

\bibitem[\protect\citeauthoryear{{Fumagalli}, {Patel}, {Franx}, {Brammer}, {van
  Dokkum}, {da Cunha}, {Kriek}, {Lundgren}, {Momcheva}, {Rix}, {Schmidt},
  {Skelton}, {Whitaker}, {Labbe} \& {Nelson}}{{Fumagalli}
  et~al.}{2012}]{Fumagalli}
{Fumagalli} M.,  {Patel} S.~G.,  {Franx} M.,  {Brammer} G.,  {van Dokkum} P.,
  {da Cunha} E.,  {Kriek} M.,  {Lundgren} B.,  {Momcheva} I.,  {Rix} H.-W.,
  {Schmidt} K.~B.,  {Skelton} R.~E.,  {Whitaker} K.~E.,  {Labbe} I.,
  {Nelson} E.,  2012, ApJL, 757, L22

\bibitem[\protect\citeauthoryear{{Garn} \& {Best}}{{Garn} \&
  {Best}}{2010}]{GarnBest}
{Garn} T.,  {Best} P.~N.,  2010, MNRAS, 409, 421

\bibitem[\protect\citeauthoryear{{Garn}, {Sobral}, {Best}, {Geach}, {Smail},
  {Cirasuolo}, {Dalton}, {Dunlop}, {McLure} \& {Farrah}}{{Garn}
  et~al.}{2010}]{Garn2010a}
{Garn} T.,  {Sobral} D.,  {Best} P.~N.,  {Geach} J.~E.,  {Smail} I.,
  {Cirasuolo} M.,  {Dalton} G.~B.,  {Dunlop} J.~S.,  {McLure} R.~J.,
  {Farrah} D.,  2010, MNRAS, 402, 2017

\bibitem[\protect\citeauthoryear{{Geach}, {Smail}, {Best}, {Kurk}, {Casali},
  {Ivison} \& {Coppin}}{{Geach} et~al.}{2008}]{G08}
{Geach} J.~E.,  {Smail} I.,  {Best} P.~N.,  {Kurk} J.,  {Casali} M.,  {Ivison}
  R.~J.,    {Coppin} K.,  2008, MNRAS, 388, 1473

\bibitem[\protect\citeauthoryear{{Geach}, {Sobral}, {Hickox}, {Wake}, {Smail},
  {Best}, {Baugh} \& {Stott}}{{Geach} et~al.}{2012}]{Geach12}
{Geach} J.~E.,  {Sobral} D.,  {Hickox} R.~C.,  {Wake} D.~A.,  {Smail} I.,
  {Best} P.~N.,  {Baugh} C.~M.,    {Stott} J.~P.,  2012, MNRAS, 426, 679

\bibitem[\protect\citeauthoryear{{Hayashi}, {Sobral}, {Best}, {Smail} \&
  {Kodama}}{{Hayashi} et~al.}{2013}]{Hayashi13}
{Hayashi} M.,  {Sobral} D.,  {Best} P.~N.,  {Smail} I.,    {Kodama} T.,  2013,
  MNRAS, 430, 1042

\bibitem[\protect\citeauthoryear{{Hickox}, {Wardlow}, {Smail}, {Myers},
  {Alexander}, {Swinbank}, {Danielson}, {Stott} \& {et al.}}{{Hickox}
  et~al.}{2012}]{Hickox}
{Hickox} R.~C.,  {Wardlow} J.~L.,  {Smail} I.,  {Myers} A.~D.,  {Alexander}
  D.~M.,  {Swinbank} A.~M.,  {Danielson} A.~L.~R.,  {Stott} J.~P.,    {et al.}
  2012, MNRAS, 421, 284

\bibitem[\protect\citeauthoryear{{Hopkins} \& {Beacom}}{{Hopkins} \&
  {Beacom}}{2006}]{Hopkins2006}
{Hopkins} A.~M.,  {Beacom} J.~F.,  2006, ApJ, 651, 142

\bibitem[\protect\citeauthoryear{{Hopkins}, {Bundy}, {Murray}, {Quataert},
  {Lauer} \& {Ma}}{{Hopkins} et~al.}{2009}]{Hopkins09}
{Hopkins} P.~F.,  {Bundy} K.,  {Murray} N.,  {Quataert} E.,  {Lauer} T.~R.,
  {Ma} C.-P.,  2009, MNRAS, 398, 898

\bibitem[\protect\citeauthoryear{{Ibar}, {Sobral}, {Best}, {Ivison}, {Smail},
  {Arumugam}, {Berta} \& {et al.}}{{Ibar} et~al.}{2013}]{Ibar13}
{Ibar} E.,  {Sobral} D.,  {Best} P.~N.,  {Ivison} R.~J.,  {Smail} I.,
  {Arumugam} V.,  {Berta} S.,    {et al.} 2013, MNRAS, 434, 3218

\bibitem[\protect\citeauthoryear{{Ilbert}, {McCracken}, {Le F{\`e}vre},
  {Capak}, {Dunlop}, {Karim}, {Renzini} \& {et al.}}{{Ilbert}
  et~al.}{2013}]{Ilbert}
{Ilbert} O.,  {McCracken} H.~J.,  {Le F{\`e}vre} O.,  {Capak} P.,  {Dunlop} J.,
   {Karim} A.,  {Renzini} M.~A.,    {et al.} 2013, A\&A, 556, A55

\bibitem[\protect\citeauthoryear{{Juneau}, {Glazebrook}, {Crampton},
  {McCarthy}, {Savaglio}, {Abraham}, {Carlberg}, {Chen}, {Le Borgne}, {Marzke},
  {Roth}, {J{\o}rgensen}, {Hook} \& {Murowinski}}{{Juneau}
  et~al.}{2005}]{Juneau}
{Juneau} S.,  {Glazebrook} K.,  {Crampton} D.,  {McCarthy} P.~J.,  {Savaglio}
  S.,  {Abraham} R.,  {Carlberg} R.~G.,  {Chen} H.-W.,  {Le Borgne} D.,
  {Marzke} R.~O.,  {Roth} K.,  {J{\o}rgensen} I.,  {Hook} I.,    {Murowinski}
  R.,  2005, ApJL, 619, L135

\bibitem[\protect\citeauthoryear{{Karim}, {Schinnerer},
  {Mart{\'{\i}}nez-Sansigre}, {Sargent}, {van der Wel}, {Rix}, {Ilbert},
  {Smol{\v c}i{\'c}}, {Carilli}, {Pannella}, {Koekemoer}, {Bell} \&
  {Salvato}}{{Karim} et~al.}{2011}]{Karim}
{Karim} A.,  {Schinnerer} E.,  {Mart{\'{\i}}nez-Sansigre} A.,  {Sargent} M.~T.,
   {van der Wel} A.,  {Rix} H.-W.,  {Ilbert} O.,  {Smol{\v c}i{\'c}} V.,
  {Carilli} C.,  {Pannella} M.,  {Koekemoer} A.~M.,  {Bell} E.~F.,    {Salvato}
  M.,  2011, ApJ, 730, 61

\bibitem[\protect\citeauthoryear{{Kennicutt}
  Jr.}{{Kennicutt}}{1998}]{Kennicutt}
{Kennicutt} Jr. R.~C.,  1998, ARAA, 36, 189

\bibitem[\protect\citeauthoryear{{Koyama}, {Smail}, {Kurk}, {Geach}, {Sobral},
  {Kodama}, {Nakata}, {Swinbank}, {Best}, {Hayashi} \& {Tadaki}}{{Koyama}
  et~al.}{2013}]{Koyama13}
{Koyama} Y.,  {Smail} I.,  {Kurk} J.,  {Geach} J.~E.,  {Sobral} D.,  {Kodama}
  T.,  {Nakata} F.,  {Swinbank} A.~M.,  {Best} P.~N.,  {Hayashi} M.,
  {Tadaki} K.-i.,  2013, MNRAS, 434, 423

\bibitem[\protect\citeauthoryear{{Lawrence}, {Warren}, {Almaini}, {Edge},
  {Hambly}, {Jameson}, {Lucas} \& {et al.}}{{Lawrence} et~al.}{2007}]{Lawrence}
{Lawrence} A.,  {Warren} S.~J.,  {Almaini} O.,  {Edge} A.~C.,  {Hambly} N.~C.,
  {Jameson} R.~F.,  {Lucas} P.,    {et al.} 2007, MNRAS, 379, 1599

\bibitem[\protect\citeauthoryear{{Leitner}}{{Leitner}}{2012}]{Leitner12}
{Leitner} S.~N.,  2012, ApJ, 745, 149

\bibitem[\protect\citeauthoryear{{Li}, {Mo} \& {Gao}}{{Li} et~al.}{2008}]{Li08}
{Li} Y.,  {Mo} H.~J.,    {Gao} L.,  2008, MNRAS, 389, 1419

\bibitem[\protect\citeauthoryear{{Lilly}, {Le Fevre}, {Hammer} \&
  {Crampton}}{{Lilly} et~al.}{1996}]{Lilly96}
{Lilly} S.~J.,  {Le Fevre} O.,  {Hammer} F.,    {Crampton} D.,  1996, ApJL,
  460, L1

\bibitem[\protect\citeauthoryear{{Ly}, {Malkan}, {Kashikawa}, {Shimasaku},
  {Doi}, {Nagao}, {Iye} \& {et al.}}{{Ly} et~al.}{2007}]{Ly2007}
{Ly} C.,  {Malkan} M.~A.,  {Kashikawa} N.,  {Shimasaku} K.,  {Doi} M.,  {Nagao}
  T.,  {Iye} M.,    {et al.} 2007, ApJ, 657, 738

\bibitem[\protect\citeauthoryear{{Magnelli}, {Elbaz}, {Chary}, {Dickinson}, {Le
  Borgne}, {Frayer} \& {Willmer}}{{Magnelli} et~al.}{2009}]{Magnelli}
{Magnelli} B.,  {Elbaz} D.,  {Chary} R.~R.,  {Dickinson} M.,  {Le Borgne} D.,
  {Frayer} D.~T.,    {Willmer} C.~N.~A.,  2009, A\&A, 496, 57

\bibitem[\protect\citeauthoryear{{Marchesini}, {van Dokkum}, {F{\"o}rster
  Schreiber}, {Franx}, {Labb{\'e}} \& {Wuyts}}{{Marchesini}
  et~al.}{2009}]{Marchesini}
{Marchesini} D.,  {van Dokkum} P.~G.,  {F{\"o}rster Schreiber} N.~M.,  {Franx}
  M.,  {Labb{\'e}} I.,    {Wuyts} S.,  2009, ApJ, 701, 1765

\bibitem[\protect\citeauthoryear{{Martin}, {Seibert}, {Buat},
  {Iglesias-P{\'a}ramo}, {Barlow}, {Bianchi} \& {et al.}}{{Martin}
  et~al.}{2005}]{Martin}
{Martin} D.~C.,  {Seibert} M.,  {Buat} V.,  {Iglesias-P{\'a}ramo} J.,  {Barlow}
  T.~A.,  {Bianchi} L.,    {et al.} 2005, ApJL, 619, L59

\bibitem[\protect\citeauthoryear{{Mobasher}, {Dahlen}, {Hopkins}, {Scoville},
  {Capak}, {Rich}, {Sanders}, {Schinnerer} \& {et al.}}{{Mobasher}
  et~al.}{2009}]{Mobasher09}
{Mobasher} B.,  {Dahlen} T.,  {Hopkins} A.,  {Scoville} N.~Z.,  {Capak} P.,
  {Rich} R.~M.,  {Sanders} D.~B.,  {Schinnerer} E.,    {et al.} 2009, ApJ, 690,
  1074

\bibitem[\protect\citeauthoryear{{Moster}, {Naab} \& {White}}{{Moster}
  et~al.}{2013}]{Moster13}
{Moster} B.~P.,  {Naab} T.,    {White} S.~D.~M.,  2013, MNRAS, 428, 3121

\bibitem[\protect\citeauthoryear{{Muzzin}, {Marchesini}, {Stefanon}, {Franx},
  {McCracken}, {Milvang-Jensen}, {Dunlop} \& {et al.}}{{Muzzin}
  et~al.}{2013}]{Muzzin13}
{Muzzin} A.,  {Marchesini} D.,  {Stefanon} M.,  {Franx} M.,  {McCracken} H.~J.,
   {Milvang-Jensen} B.,  {Dunlop} J.~S.,    {et al.} 2013, ApJ, 777, 18

\bibitem[\protect\citeauthoryear{{Neistein}, {van den Bosch} \&
  {Dekel}}{{Neistein} et~al.}{2006}]{Neistein}
{Neistein} E.,  {van den Bosch} F.~C.,    {Dekel} A.,  2006, MNRAS, 372, 933

\bibitem[\protect\citeauthoryear{{Oppenheimer}, {Dav{\'e}}, {Kere{\v s}},
  {Fardal}, {Katz}, {Kollmeier} \& {Weinberg}}{{Oppenheimer}
  et~al.}{2010}]{Oppenheimer10}
{Oppenheimer} B.~D.,  {Dav{\'e}} R.,  {Kere{\v s}} D.,  {Fardal} M.,  {Katz}
  N.,  {Kollmeier} J.~A.,    {Weinberg} D.~H.,  2010, MNRAS, 406, 2325

\bibitem[\protect\citeauthoryear{{Peng}, {Lilly}, {Kova{\v c}}, {Bolzonella},
  {Pozzetti} \& {et al.}}{{Peng} et~al.}{2010}]{Peng}
{Peng} Y.-j.,  {Lilly} S.~J.,  {Kova{\v c}} K.,  {Bolzonella} M.,  {Pozzetti}
  L.,    {et al.} 2010, ApJ, 721, 193

\bibitem[\protect\citeauthoryear{{Peng}, {Lilly}, {Renzini} \&
  {Carollo}}{{Peng} et~al.}{2012}]{Peng12}
{Peng} Y.-j.,  {Lilly} S.~J.,  {Renzini} A.,    {Carollo} M.,  2012, ApJ, 757,
  4

\bibitem[\protect\citeauthoryear{{Price}, {Kriek}, {Brammer}, {Conroy},
  {Forster Schreiber}, {Franx}, {Fumagalli} \& {et al.}}{{Price}
  et~al.}{2013}]{Price}
{Price} S.~H.,  {Kriek} M.,  {Brammer} G.~B.,  {Conroy} C.,  {Forster
  Schreiber} N.~M.,  {Franx} M.,  {Fumagalli} M.,    {et al.} 2013,
  arXiv:1310.4177

\bibitem[\protect\citeauthoryear{{Robertson}, {Ellis}, {Dunlop}, {McLure} \&
  {Stark}}{{Robertson} et~al.}{2010}]{Robertson}
{Robertson} B.~E.,  {Ellis} R.~S.,  {Dunlop} J.~S.,  {McLure} R.~J.,    {Stark}
  D.~P.,  2010, Nature, 468, 49
  
\bibitem[\protect\citeauthoryear{{Salim} \& {Lee}}{2012}]{Salim2012}
{Salim} S. \&  {Lee} J. 2012, ApJ, 758, 134

\bibitem[\protect\citeauthoryear{{Scoville}, {Aussel}, {Brusa}, {Capak},
  {Carollo}, {Elvis} \& {et al.}}{{Scoville} et~al.}{2007}]{Scoville}
{Scoville} N.,  {Aussel} H.,  {Brusa} M.,  {Capak} P.,  {Carollo} C.~M.,
  {Elvis} M.,    {et al.} 2007, ApJS, 172, 1

\bibitem[\protect\citeauthoryear{{Shioya}, {Taniguchi}, {Sasaki}, {Nagao},
  {Murayama}, {Takahashi}, {Ajiki} \& {et al.}}{{Shioya} et~al.}{2008}]{Shioya}
{Shioya} Y.,  {Taniguchi} Y.,  {Sasaki} S.~S.,  {Nagao} T.,  {Murayama} T.,
  {Takahashi} M.~I.,  {Ajiki} M.,    {et al.} 2008, ApJS, 175, 128

\bibitem[\protect\citeauthoryear{{Simpson}, {Swinbank}, {Smail}, {Alexander},
  {Brandt}, {Bertoldi}, {de Breuck}, {Chapman} \& {et al.}}{{Simpson}
  et~al.}{2013}]{Simpson13}
{Simpson} J.,  {Swinbank} M.,  {Smail} I.,  {Alexander} D.,  {Brandt} N.,
  {Bertoldi} F.,  {de Breuck} C.,  {Chapman} S.,    {et al.} 2013, ApJ,
  submitted, arXiv:1310.6363

\bibitem[\protect\citeauthoryear{{Smit}, {Bouwens}, {Franx}, {Illingworth},
  {Labb{\'e}}, {Oesch} \& {van Dokkum}}{{Smit} et~al.}{2012}]{Smit}
{Smit} R.,  {Bouwens} R.~J.,  {Franx} M.,  {Illingworth} G.~D.,  {Labb{\'e}}
  I.,  {Oesch} P.~A.,    {van Dokkum} P.~G.,  2012, ApJ, 756, 14

\bibitem[\protect\citeauthoryear{{Sobral}, {Best}, {Geach}, {Smail},
  {Cirasuolo}, {Garn}, {Dalton} \& {Kurk}}{{Sobral} et~al.}{2010}]{SOBRAL10A}
{Sobral} D.,  {Best} P.~N.,  {Geach} J.~E.,  {Smail} I.,  {Cirasuolo} M.,
  {Garn} T.,  {Dalton} G.~B.,    {Kurk} J.,  2010, MNRAS, 404, 1551

\bibitem[\protect\citeauthoryear{{Sobral}, {Best}, {Geach}, {Smail}, {Kurk},
  {Cirasuolo}, {Casali} \& {et al.}}{{Sobral} et~al.}{2009a}]{S09a}
{Sobral} D.,  {Best} P.~N.,  {Geach} J.~E.,  {Smail} I.,  {Kurk} J.,
  {Cirasuolo} M.,  {Casali} M.,    {et al.} 2009a, MNRAS, 398, 75

\bibitem[\protect\citeauthoryear{{Sobral}, {Best}, {Geach}, {Smail}, {Kurk},
  {Cirasuolo}, {Casali} \& {et al.}}{{Sobral} et~al.}{2009b}]{S09b}
{Sobral} D.,  {Best} P.~N.,  {Geach} J.~E.,  {Smail} I.,  {Kurk} J.,
  {Cirasuolo} M.,  {Casali} M.,    {et al.} 2009b, MNRAS, 398, L68

\bibitem[\protect\citeauthoryear{{Sobral}, {Best}, {Matsuda}, {Smail}, {Geach}
  \& {Cirasuolo}}{{Sobral} et~al.}{2012}]{Sobral12}
{Sobral} D.,  {Best} P.~N.,  {Matsuda} Y.,  {Smail} I.,  {Geach} J.~E.,
  {Cirasuolo} M.,  2012, MNRAS, 420, 1926

\bibitem[\protect\citeauthoryear{{Sobral}, {Best}, {Smail}, {Geach},
  {Cirasuolo}, {Garn} \& {Dalton}}{{Sobral} et~al.}{2011}]{Sobral11}
{Sobral} D.,  {Best} P.~N.,  {Smail} I.,  {Geach} J.~E.,  {Cirasuolo} M.,
  {Garn} T.,    {Dalton} G.~B.,  2011, MNRAS, 411, 675

\bibitem[\protect\citeauthoryear{{Sobral}, {Smail}, {Best}, {Geach}, {Matsuda},
  {Stott}, {Cirasuolo} \& {Kurk}}{{Sobral} et~al.}{2013}]{Sobral13}
{Sobral} D.,  {Smail} I.,  {Best} P.~N.,  {Geach} J.~E.,  {Matsuda} Y.,
  {Stott} J.~P.,  {Cirasuolo} M.,    {Kurk} J.,  2013, MNRAS, 428, 1128

\bibitem[\protect\citeauthoryear{{Sobral}, {Swinbank}, {Stott}, {Matthee},
  {Bower}, {Smail}, {Best}, {Geach} \& {Sharples}}{{Sobral}
  et~al.}{2013b}]{Sobral13b}
{Sobral} D.,  {Swinbank} A.~M.,  {Stott} J.,  {Matthee} J.,  {Bower} R.~G.,
  {Smail} I.,  {Best} P.~N.,  {Geach} J.~E.,    {Sharples} R.~M.,  2013b, ApJ,
  in press, arXiv:1310.3822

\bibitem[\protect\citeauthoryear{{Stott}, {Sobral}, {Smail}, {Bower}, {Best} \&
  {Geach}}{{Stott} et~al.}{2013a}]{Stott}
{Stott} J.~P.,  {Sobral} D.,  {Smail} I.,  {Bower} R.,  {Best} P.~N.,
  {Geach} J.~E.,  2013a, MNRAS, 430, 1158

\bibitem[\protect\citeauthoryear{{Stott}, {Sobral}, {Bower}, {Smail}, {Best},
  {Matsuda}, {Hayashi}, {Geach} \& {Kodama}}{{Stott} et~al.}{2013b}]{Stott13b}
{Stott} J.~P.,  {Sobral} D.,  {Bower} R.,  {Smail} I.,  {Best} P.~N.,
  {Matsuda} Y.,  {Hayashi} M.,  {Geach} J.~E.,    {Kodama} T.,  2013b, MNRAS, in
  press, arXiv:1309.0506

\bibitem[\protect\citeauthoryear{{Swinbank}, {Sobral}, {Smail}, {Geach},
  {Best}, {McCarthy}, {Crain} \& {Theuns}}{{Swinbank}
  et~al.}{2012a}]{Swinbank12a}
{Swinbank} A.~M.,  {Sobral} D.,  {Smail} I.,  {Geach} J.~E.,  {Best} P.~N.,
  {McCarthy} I.~G.,  {Crain} R.~A.,    {Theuns} T.,  2012a, MNRAS, 426, 935

\bibitem[\protect\citeauthoryear{{Swinbank}, {Smail}, {Sobral}, {Theuns},
  {Best} \& {Geach}}{{Swinbank} et~al.}{2012b}]{Swinbank12b}
{Swinbank} A.~M.,  {Smail} I.,  {Sobral} D.,  {Theuns} T.,  {Best} P.~N.,
  {Geach} J.~E.,  2012b, ApJ, 760, 130

\bibitem[\protect\citeauthoryear{{Swinbank}, {Simpson}, {Smail}, {Harrison},
  {Hodge}, {Karim}, {Walter}, {Alexander} \& {et al.}}{{Swinbank}
  et~al.}{2013}]{Swinbank13}
{Swinbank} M.,  {Simpson} J.,  {Smail} I.,  {Harrison} C.,  {Hodge} J.,
  {Karim} A.,  {Walter} F.,  {Alexander} D.,    {et al.} 2013, MNRAS,
  submitted, arXiv:1310.6362

\bibitem[\protect\citeauthoryear{{Villar}, {Gallego}, {P{\'e}rez-Gonz{\'a}lez},
  {Pascual}, {Noeske}, {Koo}, {Barro} \& {Zamorano}}{{Villar}
  et~al.}{2008}]{Villar08}
{Villar} V.,  {Gallego} J.,  {P{\'e}rez-Gonz{\'a}lez} P.~G.,  {Pascual} S.,
  {Noeske} K.,  {Koo} D.~C.,  {Barro} G.,    {Zamorano} J.,  2008, ApJ, 677,
  169

\bibitem[\protect\citeauthoryear{{Wuyts}, {Labb{\'e}}, {Franx}, {Rudnick}, {van
  Dokkum}, {Fazio}, {F{\"o}rster Schreiber}, {Huang}, {Moorwood}, {Rix},
  {R{\"o}ttgering} \& {van der Werf}}{{Wuyts} et~al.}{2007}]{Wuyts}
{Wuyts} S.,  {Labb{\'e}} I.,  {Franx} M.,  {Rudnick} G.,  {van Dokkum} P.~G.,
  {Fazio} G.~G.,  {F{\"o}rster Schreiber} N.~M.,  {Huang} J.,  {Moorwood}
  A.~F.~M.,  {Rix} H.-W.,  {R{\"o}ttgering} H.,    {van der Werf} P.,  2007,
  ApJ, 655, 51

\bibitem[\protect\citeauthoryear{{Zheng}, {Bell}, {Papovich}, {Wolf},
  {Meisenheimer}, {Rix}, {Rieke} \& {Somerville}}{{Zheng} et~al.}{2007}]{Zheng}
{Zheng} X.~Z.,  {Bell} E.~F.,  {Papovich} C.,  {Wolf} C.,  {Meisenheimer} K.,
  {Rix} H.-W.,  {Rieke} G.~H.,    {Somerville} R.,  2007, ApJL, 661, L41

\end{thebibliography}

\appendix

\section[]{SFR Functions for different stellar mass bins} \label{reliability_und}

%
%
\begin{table*}
 \centering
  \caption{The SFR function and its evolution for $0.40<z<2.23$ for samples of star-forming galaxies with different stellar masses. The samples of star-forming galaxies are divided in 3 bins of stellar masses (dex) per redshift: $8.7-9.3$, $9.3-10.0$ and $10.0-11.5$. SFR functions are fitted with a Schechter function and further constrained by the appropriate stellar mass function: we require that the number density in the full SFR for a given mass bin is consistent (within the errors) with the number density of galaxies implied from the stellar mass function. In order to constrain $\alpha$, we search for the best common value for each mass bin. All mass bins present a common $\alpha$ within less than 1\,$\sigma$ of the best fit and that comply to the physical conditions applied. $\alpha$ is thus fixed at these values. We note that SFR$^*$ values increase (and $\Phi*$ values decrease) for steeper $\alpha$ and decrease for shallower $\alpha$, but they vary consistently for all masses, and thus the differences between masses are maintained, within the errors. FC $\rho_{\rm SFR,obsM} $ presents the fractional contribution to $\rho_{\rm SFR,obsM}$ (SFGs with masses $>10^{8.7}$\,M$_{\odot}$), while FC $\rho_{\rm SFR,Tot} $ is the fractional contribution to the total $\rho_{\rm SFR} $.}
  \begin{tabular}{@{}cccccccc@{}}
  \hline
   \bf Redshift & $\log_{10}$M & $\bf SFR^*$ & $\rm \bf \Phi^*$ & $\alpha$ & $\rho_{\rm SFR} $  & FC $\rho_{\rm SFR,obsM} $ & FC $\rho_{\rm SFR,Tot} $   \\
     ($z$)    & (M$_{\odot}$) & (M$_{\odot}$\,yr$^{-1}$) & (Mpc$^{-3}$) &  & (M$_{\odot}$\,yr$^{-1}$ Mpc$^{-3}$) & (\%) & (\%)  \\
 \hline
   \noalign{\smallskip}
$0.40$ & $ 8.7-9.3$ &  $0.8^{+  0.1}_{-  0.1}$ & $-2.87^{+ 0.03}_{- 0.07}$ & 0.2 & $0.0022\pm0.0011$ & $24\pm11$ & $18\pm9$ \\ 
 & $ 9.3-10.0$ & $  4^{+ 8}_{-  2}$ & $-3.48^{+ 0.17}_{- 0.26}$ & 0.35 & $0.0028\pm0.0009$ & $30\pm10$ & $23\pm8$  \\
 & $10.0-11.5$ & $  6.7^{+  3.6}_{-  1.6}$ & $-3.53^{+ 0.15}_{- 0.17}$ & 0.5 & $0.0042\pm0.0011$  & $46\pm11$ & $35\pm9$  \\
 \hline
$0.84$ & $ 8.7-9.3$ & $  1.7^{+  0.2}_{-  0.1}$ & $-2.53^{+ 0.03}_{- 0.03}$ & 0.2 & $0.0075\pm0.0012$  & $23\pm4$  & $17\pm3$  \\ 
 & $ 9.3-10.0$ & $  6.0^{+  1.4}_{-  0.8}$ & $-2.79^{+ 0.10}_{- 0.11}$ & 0.35 & $0.0103\pm0.0043$ & $33\pm14$  & $23\pm10$  \\ 
 & $10.0-11.5$ & $ 11.7^{+  1.9}_{-  1.5}$ & $-2.91^{+ 0.07}_{- 0.08}$ & 0.5 & $0.014\pm0.007$ & $44\pm22$  & $31\pm16$  \\  
 \hline
$1.47$ & $ 8.7-9.3$ & $  3.0^{+  0.2}_{-  0.2}$ & $-2.19^{+ 0.03}_{- 0.07}$ & 0.2 & $0.010\pm0.005$  & $20\pm10$ & $11\pm6$  \\ 
 & $9.3-10.0$ & $ 11.6^{+  1.2}_{-  1.7}$ & $-2.65^{+ 0.12}_{- 0.12}$ & 0.35 & $0.0144\pm0.005$   & $29\pm10$ & $15\pm6$  \\ 
 & $10.0-11.5$ & $ 28.6^{+  1.4}_{-  1.3}$ & $-2.97^{+ 0.06}_{- 0.06}$ & 0.5 & $0.025\pm0.008$ & $51\pm16$ & $26\pm9$  \\ 
 \hline
$2.23$ & $ 8.7-9.3$ &  $  3.5^{+  0.1}_{-  0.1}$ & $-2.26^{+ 0.10}_{- 0.03}$ & 0.2 & $0.011\pm0.004$  & $13\pm5$ & $10\pm4$  \\  
 & $ 9.3-10.0$ & $ 17.6^{+  2.1}_{-  1.7}$ & $-2.63^{+ 0.06}_{- 0.08}$ & 0.35 & $0.02\pm0.009$ & $23\pm10$ & $17\pm8$  \\ 
 & $10.0-11.5$ & $ 64^{+ 9}_{- 8}$ & $-3.23^{+ 0.08}_{- 0.08}$ & 0.5 & $0.057\pm0.019$ & $65\pm22$ & $49\pm17$  \\ 

 \hline
\end{tabular}
\label{SFRD_diff_MASSES}
\end{table*}

\label{lastpage}

\end{document}